\documentclass[aps,prb,reprint,groupedaddress,nofootinbib]{revtex4-2}

\usepackage{braket}
\usepackage{graphicx}
\usepackage{amsmath}
\usepackage{siunitx}
\usepackage{makecell} 

\usepackage[math,column=X]{cellspace}
\graphicspath{{figures/}}

\newcommand{\centeredgraphics}[1]{\noindent\makebox[\columnwidth]{\includegraphics{#1}}} 

\usepackage{textcomp}
\newcommand{\tsim}{\raisebox{0.5ex}{\texttildelow}}

\renewcommand{\Im}{\,\text{Im}}
\renewcommand{\tensor}[1]{\bar{\mathbf{#1}}}

\newcommand{\aalpha}{\boldsymbol{\alpha}}
\newcommand{\ddelta}{\boldsymbol{\delta}}
\newcommand{\eepsilon}{\boldsymbol{\epsilon}}
\newcommand{\E}{\mathcal{E}}
\newcommand{\EE}{\boldsymbol{\E}}
\newcommand{\gradk}{\boldsymbol\nabla_\kk}
\newcommand{\berry}{\Omega^z_n}
\newcommand{\bberry}{\boldsymbol{\Omega}_n}
\newcommand{\m}{\mu_n^z}
\newcommand{\mm}{\boldsymbol{\mu}_n}
\newcommand{\M}{\mathcal{M}}
\newcommand{\MM}{\boldsymbol{\M}}
\newcommand{\kk}{\mathbf{k}} 
\newcommand{\dkx}{\partial_{k_x}}
\newcommand{\dky}{\partial_{k_y}}

\DeclareSIUnit\muB{\mbox{$\mu_{\text{B}}$}}
\DeclareSIUnit[per-mode=symbol]{\magperum}{\muB\per\square\micro\meter}
\DeclareSIUnit[per-mode=reciprocal]{\magperumHz}{\magperum\per\Hz\tothe{1/2}}
\DeclareSIUnit{\micron}{\micro\metre}

\newcommand \dd[1]  { \,\textrm d{#1} }   

\renewcommand{\eqref}[1]{Eq.~(\ref{eq:#1})}

\newcommand{\tabref}[1]{Table~\ref{tab:#1}}
\newcommand{\secref}[1]{Sec.~\ref{sec:#1}}

\newcommand{\appref}[1]{Appendix~\ref{app:#1}}
\newcommand{\figref}[2][]{Fig.~\ref{fig:#2}%
    \ifx&#1&%
    \else
    (#1)
    \fi} 
\newcommand{\Figref}[2][]{Figure~\ref{fig:#2}%
    \ifx&#1&%
    \else
    (#1)
    \fi}     

\begin{document}

\title{Electrically tunable and reversible magnetoelectric coupling in strained bilayer graphene}

\author{Brian~T.~Schaefer}\affiliation{Laboratory of Atomic and Solid-State Physics, Cornell University, Ithaca, NY 14853, USA}
\author{Katja~C.~Nowack}\affiliation{Laboratory of Atomic and Solid-State Physics, Cornell University, Ithaca, NY 14853, USA}\affiliation{Kavli Institute at Cornell for Nanoscale Science, Cornell University, Ithaca, NY 14853, USA}\email{kcn34@cornell.edu}

\date{\today}

\begin{abstract}
 The valleys in hexagonal two-dimensional systems with broken inversion symmetry carry an intrinsic orbital magnetic moment. 
 Despite this, such systems possess zero net magnetization unless additional symmetries are broken, since the contributions from both valleys cancel. 
 A nonzero net magnetization can be induced through applying both uniaxial strain to break the rotational symmetry of the lattice and an in-plane electric field to break time-reversal symmetry owing to the resulting current. 
 This creates a magnetoelectric effect whose strength is characterized by a magnetoelectric susceptibility, which describes the induced magnetization per unit applied in-plane electric field.
 Here, we predict the strength of this magnetoelectric susceptibility for Bernal-stacked bilayer graphene as a function of the magnitude and direction of strain, the chemical potential, and the interlayer electric field. 
 We estimate that an orbital magnetization of \tsim\SI{5400}{\magperum} can be achieved for \SI{1}{\percent} uniaxial strain and a \SI{10}{\micro\ampere} bias current, which is almost three orders of magnitude larger than previously probed experimentally in strained monolayer MoS$_2$. 
 We also identify regimes in which the magnetoelectric susceptibility not only switches sign upon reversal of the interlayer electric field but also in response to small changes in the carrier density. 
 Taking advantage of this reversibility, we further show that it is experimentally feasible to probe the effect using scanning magnetometry.
\end{abstract}

\maketitle

\section{Introduction}

Two-dimensional hexagonal Dirac materials are a promising platform for realizing orbital magnetic effects.
In these materials, the low-energy band structure features two degenerate energy minima (or ``valleys'') at the K and K$'$ points at the corners of the Brillouin zone~\cite{theoryreview}.
If inversion symmetry is broken, the energy bands are directly gapped at each valley.
In this case, the electronic states in each valley are characterized by a strong intrinsic orbital magnetic moment and Berry curvature~\cite{berryreview}.
These quantities differ in sign between the two valleys, exhibiting distributions centered at K and K$'$ with maxima that typically increase with decreasing magnitude of the gap~\cite{valleycontrasting}.
The control of orbital magnetic moments is both fundamentally and technologically interesting: it provides a direct window into phenomena driven by the Berry curvature and may provide an efficient way to switch magnetic layers through the generation of strong magnetic torques~\cite{2dspintronicreview, faijiereview}.
However, in equilibrium the total magnetization, which depends on both the orbital magnetic moment and Berry curvature, is precisely zero due to equal and opposite contributions from each valley.
Different strategies to induce and detect net magnetic moments have been demonstrated previously in transition metal dichalcogenide (TMD) devices~\cite{valleyhall, valleyhalljieun, lizhong, mos2pull, mos2bend}.

Here, we focus on Bernal-stacked bilayer graphene (BLG) which is promising for generating a strong, purely orbital magnetization for several reasons.
First, the maximum orbital magnetic moment and Berry curvature in each valley are expected to be inversely related to the interlayer asymmetry $\Delta$ (see \appref{twoband})~\cite{valleycontrasting}.
This quantity describes a potential energy difference between the two layers and controls the size of the bandgap.
In dual-gated BLG devices, the size and sign of $\Delta$ can be tuned independently of the carrier density through an interlayer electric field~\cite{BLGgap, valleycurrentBLG, valleycurrentBLG2}.
The low charge inhomogeneity in state-of-the-art graphene-based devices enables operation at low carrier densities~\cite{vdwYankowitz}, which is necessary to take advantage of the enhanced magnetic moment and Berry curvature at small bandgap.
Second, graphene has a nearly vanishing spin-orbit coupling~\cite{theoryreview}.
This suggests that the magnetization in BLG is entirely orbital in origin, in contrast with TMDs in which spin contributions can be intertwined with orbital effects~\cite{berryreview, spinorbitTMD}.
Finally, BLG has a rich low-energy fermiology due to trigonal warping of the band structure, offering an interesting platform in which to study orbital magnetism~\cite{mccann, moulsdale, berrydipole}.

Naturally, one route towards generating a net orbital magnetization is to create a net valley polarization~\cite{valleycontrasting}.
This can be achieved by selective optical excitation of electrons in a single valley using circularly polarized light as demonstrated previously in MoS$_2$~\cite{valleyhall, valleyhalljieun}. 
An electrically tunable valley polarization has also been realized in WSe$_2$/CrI$_3$ heterostructures, in which the valley polarization of WSe$_2$ is controlled through proximity coupling to CrI$_3$~\cite{lizhong}.  
However, the terahertz-scale optical transitions and lack of spin-orbit coupling in BLG make these methods difficult to extend to BLG.

An alternative way to create a net orbital magnetization was demonstrated in uniaxially strained MoS$_2$ devices through a magnetoelectric effect: application of an in-plane electric field drives a transport current that induces a net orbital magnetization~\cite{mos2pull, mos2bend}.
Here, we consider the analogous effect in strained bilayer graphene (sBLG).
This effect does not rely on a net valley polarization but rather on the combination of strain, which breaks the rotational symmetry of the lattice, a bias current, which breaks time-reversal symmetry, and an interlayer electric field, which breaks layer inversion symmetry.

\begin{figure}
\centeredgraphics{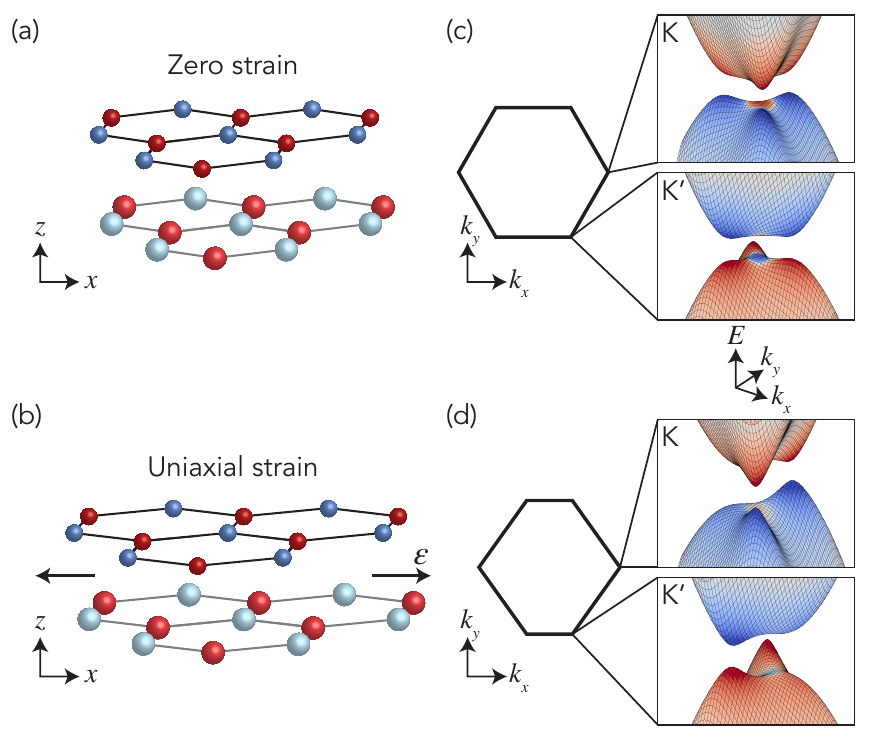}
\caption{
(a)-(b) Schematics of the BLG lattice under (a) zero strain and (b) uniaxial tensile strain along the $x$ zigzag crystal axis.
(c)-(d) Brillouin zone (geometry exaggerated) and low-energy band structure for the K and K$'$ valleys calculated from the model below (see \secref{model}) with $\Delta=7$~meV, $\theta=0$, and (c) $\varepsilon=0$ or (d) $\varepsilon=0.01$.
The intensity (color) of the shading represents the magnitude (sign) of the quantity $\MM$, which captures contributions from the orbital magnetic moment and Berry curvature.
$\MM$ and the color scale are more precisely defined in \secref{linearME} and \figref{susc} below.
}
\label{fig:intro}
\end{figure}

In the following, we briefly illustrate the effect using results from the model described in \secref{theory}.
\Figref[a]{intro} shows the lattice, Brillouin zone, and low-energy band structure for gapped BLG at zero strain.
The colored shading on the energy bands represents the magnitude and sign of the quantity $\MM$ as defined below. 
This quantity captures the contributions of occupied electronic states to the total magnetization from both the orbital magnetic moment and the Berry curvature.
Close to the K and K$'$ points, trigonal warping of the band structure gives rise to three mini-valleys which result in  hotspots of $\MM$~\cite{moulsdale, berrydipole}.
Applying uniaxial strain to the BLG lattice breaks the three-fold rotational symmetry and distorts the energy bands and magnetic moment distribution as shown in \figref[b]{intro}.
Despite this distortion, the distributions of $\MM$ in the two valleys are still equal in magnitude and opposite in sign, leading to zero net magnetization. 
However, an in-plane electric field creates an electric current that breaks time-reversal symmetry. 
As a result, the electronic states contributing to the net magnetization are described by non-equilibrium occupation functions which are shifted in the same direction in momentum space for each valley. 
Integrating over contributions from occupied states in each valley therefore leads to a net bulk magnetization that is purely orbital in nature.
The strength of this effect is characterized by a magnetoelectric susceptibility, i.e., the coefficient describing the magnitude of induced magnetization per unit applied electric field.
The sign of the magnetoelectric susceptibility can be switched by reversal of either the in-plane or the interlayer electric field. 

The Berry curvature dipole, which is related to the magnetoelectric susceptibility, has been previously studied in hexagonal Dirac materials in the context of nonlinear transport~\cite{berrywte2, berrywte2b, kaifei, sodemann, berrydipole}. 
In particular, Battilomo \emph{et al}. show that interlayer hopping processes that induce trigonal warping lead to a finite Berry curvature dipole in sBLG that can exhibit sign reversal upon continuous tuning of the carrier density~\cite{berrydipole}. 
Here, we also find regimes where the sign changes in response to small changes in the carrier density. 
These reversals are associated with changes in the topology of the Fermi surface such as the formation of an additional Fermi surface pocket or merging of pockets.

Recently, strong orbital magnetic effects have also been discovered and explored both experimentally and theoretically in twisted bilayer graphene (TBG)~\cite{dgg, orbitalferromagnetism, orbmagEfetov, he, he2, strainTBG, sotmoire, moirestm}.
The magnetization in TBG can be switched electrically in some regimes through small changes in either carrier density or an applied bias current. 
Recent work suggests that the latter can be explained by a magnetoelectric effect similar to the one considered here~\cite{he, he2}.

This paper is organized as follows. 
In \secref{model}, we describe the tight-binding model used to calculate the energy bands and eigenstates for sBLG under uniaxial strain.
Then, in \secref{linearME}, we combine the orbital magnetic moment and Berry curvature to arrive at an expression for the net orbital magnetization and extract the linear magnetoelectric susceptibility.
We calculate\footnote{
Source code available at \url{https://github.com/nowacklab/blg_strain}.
} the susceptibility as a function of various model parameters in \secref{results} and discuss how its magnitude and sign can be tuned.
We estimate the magnitude of the effect in \secref{magnitude}, propose an experiment to detect the effect using scanning magnetometry in \secref{experiment}, and conclude in \secref{conclusion}.


\section{Theory}\label{sec:theory}
\subsection{Tight-binding model}\label{sec:model}

\begin{figure}
\centeredgraphics{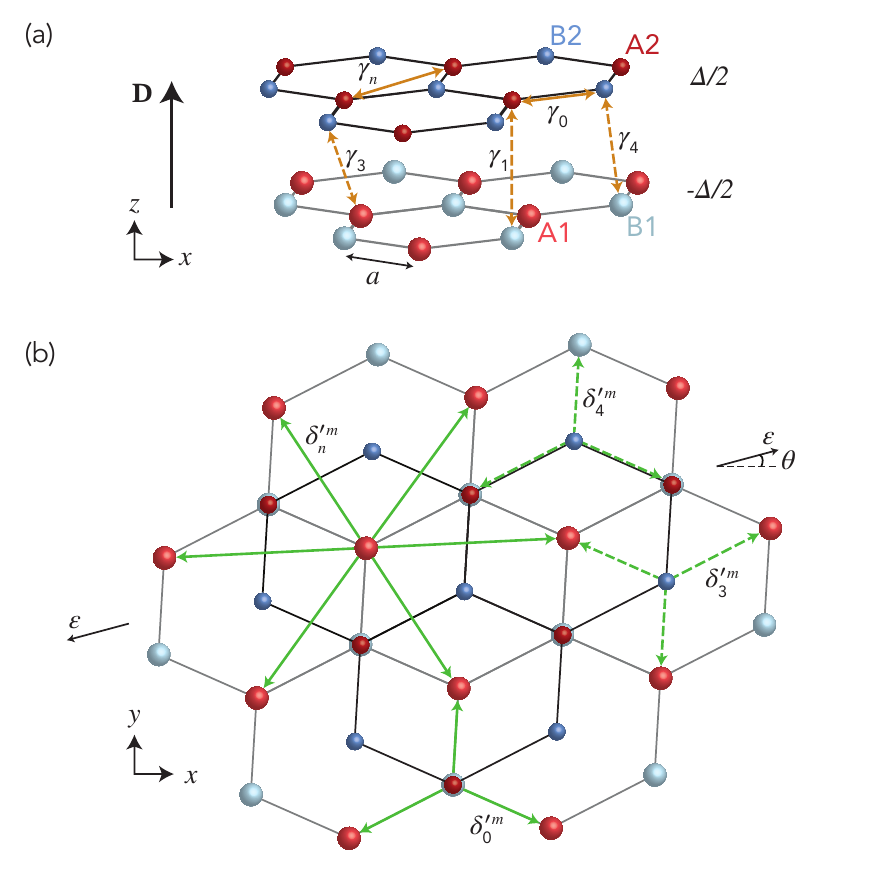}
\caption[Tight-binding model: bond vectors and hopping parameters]{
(a) Schematic of the unstrained BLG lattice, with hopping parameters $\gamma_j$, interlayer asymmetry $\Delta$, and displacement field $D$.
Atoms A2 and B1 are stacked directly on top of one another.
(b) Schematic top view of BLG lattice under uniaxial strain with magnitude $\varepsilon$ applied at an angle $\theta\approx\SI{15}{\degree}$ to the $x$ axis, with modified bond vectors ${\ddelta'}_j^m$.
The dashed (solid) arrows in each panel represent interlayer (intralayer) coupling.
}
\label{fig:tb}
\end{figure}
 
\begin{table}
\renewcommand{\arraystretch}{1.5} 
\begin{center}
\begin{tabular}{| Xc | Xc | Xc | Xc | Xc |}
 \hline
 \makecell{Hopping \\ processes} & \makecell{Matrix\\ element}          & $\gamma_j$ (eV)  & $\eta_j$  &  \makecell{Zero-strain\\ bond vectors\\ $\ddelta_j^m/a$}  \\
 \hline
 \makecell{A1-B1\\ A2-B2}  &  $h_0$         & $3.16$ & $-2$  & \makecell{$\left(\frac{\sqrt{3}}2, -\frac12\right)$;  $\left(-\frac{\sqrt{3}}2, -\frac12\right)$; \\ $(0, 1)$}   \\[5mm]
 A2-B1   & $h_1$         & $-0.381$ & --- &  $(0,0)$ \\[5mm]
 A1-B2 & $h_3$         & 0.38   & $-1$ &  \makecell{$\left(-\frac{\sqrt{3}}2, \frac12\right)$; $\left(\frac{\sqrt{3}}2, \frac12\right)$; \\ $(0, -1)$}  \\[5mm]
 \makecell{A1-A2\\ B1-B2}  & $h_4$         & 0.14  &  $-1$  &   \makecell{$\left(\frac{\sqrt{3}}2, -\frac12\right)$;  $\left(-\frac{\sqrt{3}}2, -\frac12\right)$; \\$(0, 1)$ }  \\[5mm]
\makecell{A1-A1\footnotemark[1]\\ A2-A2\footnotemark[1]\\ B1-B1\footnotemark[1]\\ B2-B2\footnotemark[1]}  &  $h_n$ & $\sim0.3$ & $-1$ &  \makecell{$\left(\sqrt{3}, 0\right)$; $\left(-\sqrt{3}, 0\right)$; \\  $\left(\frac{\sqrt{3}}2, \frac32\right)$; $\left(-\frac{\sqrt{3}}2, \frac32\right)$; \\ $\left(\frac{\sqrt{3}}2, -\frac32\right)$;  $\left(-\frac{\sqrt{3}}2, -\frac32\right)$ } \\
\hline
\end{tabular}
\end{center}
\footnotetext[1]{next-nearest neighbor}

\caption[Tight-binding model hopping parameters]{
Hopping processes in the tight-binding model and corresponding matrix elements $h_j$, Slonczewski-Weiss-McClure hopping parameters $\gamma_j$, Gr\"uneisen parameters $\eta_j$, and zero-strain bond vectors $\ddelta_j^m$ reported in units of the carbon-carbon distance $a=0.142$~nm.
The magnitudes of the hopping parameters follow the values reported in Ref.~\cite{hoppingparams} and the typical estimate $\gamma_n\sim0.1\gamma_0$~\cite{peres}.
The signs of the hopping parameters are chosen to address the ambiguity discussed in Ref.~\cite{tbsign, warpingorientation}.
The magnitude of the Gr\"uneisen parameter $\eta_0$ follows from both Raman spectroscopy measurements ($\eta_0\approx-1.99$) and first principles calculations ($\eta_0\approx-1.87$)~\cite{gruneisen}.
The estimate $\eta_{3,4,n}\sim-1$ accounts for the longer intralayer and next-nearest neighbor bond lengths~\cite{moulsdale, gaugefields}.
}
\label{tab:hopping}
\end{table}

Based on the Hamiltonian for unstrained BLG~\cite{theoryreview, interlayerasymmetry, mccann}, we construct a tight-binding model for sBLG with a $4\times4$ Hamiltonian yielding four energy bands $E_n$ labeled with $n\in(0,1,2,3)$ from lowest to highest energy.
The wavefunction $\boldsymbol{\Psi}_n(\kk)$ for each band has components $\psi_n^{\sigma i}(\kk)$ corresponding to the wavefunction amplitude for each layer $\sigma\in(\text{A},\text{B})$ and sublattice $i\in(1,2)$.
Written in the (A1, B2, A2, B1) basis, the Hamiltonian and its eigenstates are
\renewcommand{\arraystretch}{1.2}
$$
H = \left(
\begin{array}{cccc}
    -\frac12\Delta &   h_3   &   h_4  &   h_0   \\
    h_3^*     &   \frac12\Delta & h_0^*   &   h_4^*   \\
    h_4^*       &   h_0       &   \frac12\Delta + \Delta'  &   h_1   \\
    h_0^*       &   h_4  &   h_1^*   &   -\frac12\Delta + \Delta' \\
\end{array}
\right) + h_n \tensor{I}_4
$$
$$
\ket{n}\equiv\boldsymbol{\Psi}_n(\kk)=
\left(\begin{array}{c}
\psi_{n}^{\text{A1}}(\kk)\\
\psi_{n}^{\text{B2}}(\kk)\\
\psi_{n}^{\text{A2}}(\kk)\\
\psi_{n}^{\text{B1}}(\kk)
\end{array}\right),
$$
where $\tensor{I}_4$ is the $4\times4$ identity matrix and the elements $h_j$ are defined below. 
$\Delta$ is the interlayer asymmetry induced by an applied electric displacement field $D$ between the layers (\figref[a]{tb}) and $\Delta'\sim0.022$~eV accounts for a small energy cost associated with the dimerization of B1-A2 atoms~\cite{moulsdale, hoppingparams, blgtopicalreview}.

The matrix elements $h_j$ describe inter- and intralayer interactions using the Slonczewski-Weiss-McClure parameterization (\tabref{hopping} and \figref[a]{tb})~\cite{Slonczewski, McClure}.
Each $h_j$ is the product of a hopping parameter $\gamma_j$ and a structure factor that depends on  the relevant bond vectors ${\ddelta}_j^m$ (\tabref{hopping} and \figref[a]{tb}).
The subscript $j$ denotes either intralayer nearest neighbor ($j=0$), dimer ($j=1$), interlayer ($j=3,4$), or intralayer next-nearest-neighbor ($j=``n"$) interactions.
Application of strain leads to modified bond vectors ${\ddelta'}_j^m$ that depend on the strength and the orientation of the strain. 
The changes in bond lengths directly modify the hopping parameters $\gamma_j^m$ as well as the structure factors.
In total, this can be captured by matrix elements with the more general form $
h_j=\sum_m \gamma^m_j e^{i\mathbf{k}\cdot{\ddelta'_j}^m},
$ 
where the index $m$ runs over the bonds listed in \tabref{hopping}.

To linear order the modified bond vectors are given by
$$
{\ddelta'_j}^{m} = \left(\tensor{I}_2+\bar{\eepsilon}\right) \cdot {\ddelta}_j^m,
$$
where $\tensor{I}_2$ is the $2\times2$ identity matrix and $\bar{\eepsilon}$ is an arbitrary two-dimensional strain tensor~\cite{tbstrain, strainreview}.
The corresponding hopping parameter is expected to depend exponentially on changes in the bond length following
$$
\gamma_j^m = \gamma_j e^{\eta_j\left(\left|{\ddelta'_j}^{m}\right|/\left|\ddelta_j^m\right|-1\right)},
$$
where 
$\eta_j$
is the appropriate Gr\"uneisen parameter (\tabref{hopping})~\cite{tbstrain, revisited, strainreview}.
For uniaxial tensile strain\footnote{Here, we focus on tensile strain because graphene-based devices often possess a small critical buckling strain in compression~\cite{buckling, buckling3}.} as illustrated in \figref[b]{tb}, the strain tensor is~\cite{tbstrain}
\begin{equation}
\bar{\eepsilon} = \varepsilon\left(
\begin{array}{cc}
\cos^2\theta - \nu\sin^2\theta & (1+\nu)\cos\theta\sin\theta\\
(1+\nu)\cos\theta\sin\theta & \sin^2\theta - \nu\cos^2\theta
\end{array}
\right).
\label{eq:strain}
\end{equation}
Here, $\varepsilon$ is the strain magnitude, $\theta$ is the angle between the principal strain axis and the $x$ axis, and $\nu$ is the Poisson's ratio.
Following \figref{tb}, $\theta=0$ ($\theta=\pi/2$) corresponds to strain along a zigzag (armchair) axis of the crystal.
We use $\nu\approx0.165$ as the Poisson's ratio for graphene~\cite{tbstrain}, but generically if strain is transferred via adhesion to a flexible substrate, the relevant Poisson's ratio is that of the substrate~\cite{gruneisen}.

The Hamiltonian explicitly depends on the applied strain $\bar{\eepsilon}(\varepsilon,\theta)$ and the interlayer asymmetry $\Delta$ (\figref[a]{tb}). 
Below we diagonalize the Hamiltonian for each combination of $\bar{\eepsilon}$ and $\Delta$ to obtain the energy bands and eigenstates over a momentum-space grid around the K valley. 
We obtain the energy bands and eigenstates at the K$'$ valley by using the symmetry of the Hamiltonian $H(k_x, k_y) = H(-k_x, -k_y)$, valid even under uniaxial strain.

\subsection{Linear magnetoelectric susceptibility}\label{sec:linearME}\label{sec:mag}

\begin{figure*}
\centeredgraphics{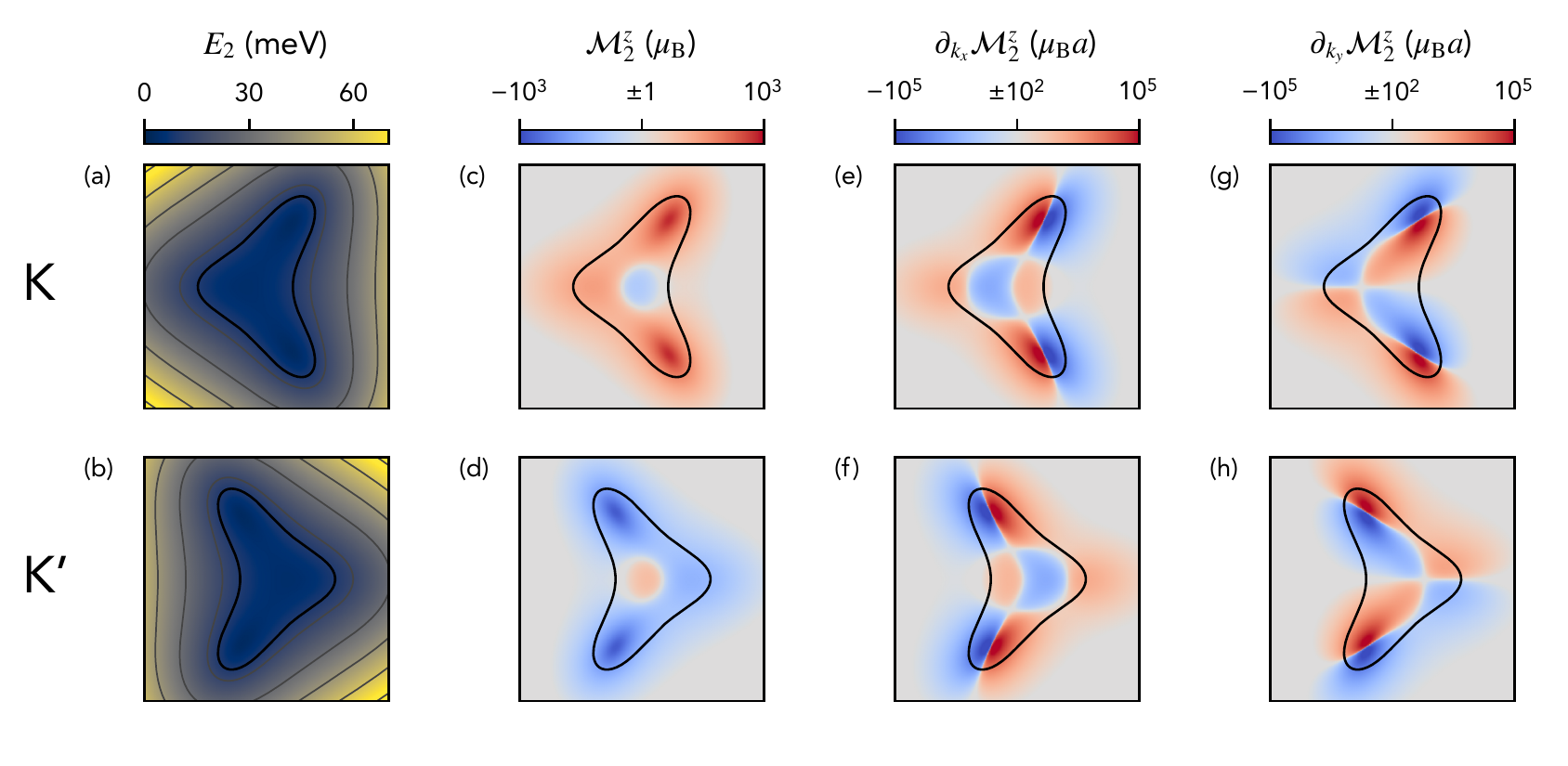}
\caption{
(a)-(b) Conduction band, (c)-(d) $\MM$ distributions, and (e)-(h) their gradients for sBLG in the K (top row) and K$'$ (bottom row) valley.
The black contour outlines the Fermi surface at chemical potential $\mu=10$~meV.
The model parameters are $\Delta=7$~meV, $\varepsilon=0.01$, and $\theta=\SI{0}{\degree}$ (strain applied along the $x$ zigzag axis).
The maps span a $0.05a^{-1}\times0.05a^{-1}$ region of momentum space centered at each valley.
Panels (c)-(h) use a logarithmic color scale, where the neutral-colored regions represent regions of momentum space in which $|\M_2^z| \leq \mu_{\text{B}}$ or $|\partial_{k_x}\M_2^z| \leq 10^2$~$\mu_{\text{B}}a$.
}
\label{fig:kspace}
\end{figure*}

The orbital magnetization includes contributions from the orbital magnetic moment $\mm$ and Berry curvature $\bberry$ for band $n$. 
These can be calculated using the standard expressions~\cite{berryreview, moulsdale}:
\begin{widetext}
\begin{equation*}
\begin{gathered}
\mm(\kk) = \frac{e}{2\hbar}i\bra{\gradk n}\times\left[E_n(\kk)-H(\kk)\right]\ket{\gradk n} = -\frac{e}{\hbar} \Im \sum_{m\neq n} \frac{\braket{n | \dkx  H | m} \braket{m | \dky H | n}}{E_n(\kk) - E_m(\kk)}\hat{\mathbf z}\\
\bberry(\kk) = i\bra{\gradk n}\times\ket{\gradk n} = -2 \Im \sum_{m\neq n} \frac{\braket{n | \dkx  H | m} \braket{m | \dky H | n}}{\left[E_n(\kk) - E_m(\kk)\right]^2}\hat{\mathbf z}.
\end{gathered}
\end{equation*}
Here, $\ket{n}$ is the eigenstate for band $n$ and $\ket{\gradk n}$ is its momentum-space gradient.
We define the quantity $\MM_n$ to capture contributions from both the orbital magnetic moment and Berry curvature:
\begin{align}\label{eq:Mn}
\begin{split}
    \MM_n(\kk) &= \mm(\kk) + \frac{e\bberry(\kk)}{\hbar}[\mu-E_n(\kk)] \\
    &= -\frac{e}{\hbar}\Im\sum_{m\neq n} \frac{\braket{n | \dkx  H | m} \braket{m | \dky H | n}}{E_n(\kk) - E_m(\kk)} \left[
1 + 2 \frac{\mu-E_n(\kk)}{E_n(\kk) - E_m(\kk)}\right]\hat{\mathbf z}.
\end{split}
\end{align}
\end{widetext}
For the case of a two-band model with electron-hole symmetry, this expression can be further simplified (see \appref{twoband})~\cite{valleycontrasting}. 
However, we choose to work with the full four-band Hamiltonian as introduced in the previous section, with electron-hole asymmetry from the parameters $\Delta'$, $\gamma_4$, and $\gamma_n$~\cite{moulsdale, blgtopicalreview, mccann}. 
At zero temperature, the net orbital magnetization is given by an integral over all occupied states in momentum space~\cite{berrycorrection, berryreview, thonhauser, orbitalmagderivation}:
\begin{equation}
\mathbf{M} = \sum_n \int \frac{\dd{}^2 \kk}{(2\pi)^2}f_n(\kk) \MM_n(\kk).
\label{eq:mag}
\end{equation}
Here, $f_n(\kk)$ is the occupation function, which is a step function in equilibrium at zero temperature:
$$
f^0_n(\kk) = \Theta[\mu - E_n(\kk)].
$$

\Figref[a-d]{kspace} shows an example of the conduction band $E_2(\kk)$ and distribution of $\M_n^z(\kk)$ in the K and K$'$ valleys under applied strain.
The sign of $\M_n^z$ differs between the two valleys as expected~\cite{valleycontrasting}.
Strain breaks the three-fold rotational symmetry of the lattice, leading to an asymmetric distribution of $\M_n^z$ within each valley.
In equilibrium, the net magnetization (\eqref{mag}) vanishes due to equal and opposite contributions from the two valleys.
However, application of an in-plane electric field $\EE = (\E_x, \E_y)$ leads to a non-equilibrium occupation function, which to lowest order corresponds to a shift of the Fermi surface in the direction of $\EE$ for both valleys. 
As a result, the occupied states lead to a net magnetization.

Under the linear relaxation-time approximation, the occupation function can be approximated as~\cite{ashcroftmermin}
$$
f_n(\kk) \approx f_n^0(\kk) + \frac{e\tau\EE}{\hbar}\cdot\gradk f^0_n(\kk),
$$
where $\tau$ is a mean scattering time.
Inserting $f_n(\kk)$ into \eqref{mag} gives for the net orbital magnetization 
\begin{equation}\label{eq:Mz_noneq}
M_z = \frac{e\tau\EE}{\hbar} \cdot \sum_n \int \frac{\dd{}^2 \kk}{(2\pi)^2} \gradk f^0_n(\kk) \M^z_n(\kk),
\end{equation}
where the equilibrium term integrates to zero considering both valleys.
The linear relationship between the applied electric field $\EE$ and net magnetization $M_z$ describes a magnetoelectric effect.
We define the dimensionless linear magnetoelectric susceptibility $\aalpha=(\alpha_x, \alpha_y)$ such that 
\begin{equation}\label{eq:ME}
\mu_0 M_z = \tau(\aalpha \cdot \EE).
\end{equation}
Integrating \eqref{Mz_noneq} by parts and discarding the boundary term which evaluates to zero finally results in
\begin{equation*}\label{eq:susc}
\aalpha 
= -\frac{e\mu_0}{\hbar} \sum_n \int \frac{\dd{}^2 \kk}{(2\pi)^2} f^0_n(\kk) \gradk \M^z_n(\kk),
\end{equation*}
i.e., an integral of $\gradk \M^z_n(\kk)$ over occupied states.

In \figref[e-h]{kspace} we show an example of the $k_x$ and $k_y$ components of $\gradk \M^z_2(\kk)$. 
The component $\alpha_x$ ($\alpha_y$) of the magnetoelectric susceptibility is proportional to the sum of the integral over occupied states within the black contours in \figref[e,f]{kspace} (\figref[g,h]{kspace}) considering both valleys.
For the case of strain along the $x$ (zigzag, $\theta=0$) axis shown here, the contributions to $\alpha_x$ from each valley are equal in both sign and magnitude, whereas the contributions to $\alpha_y$ are zero in both valleys.
Similarly, strain along the $y$ (armchair, $\theta=\pi/2$) axis also yields zero $\alpha_y$.
In both cases, the strain tensor in \eqref{strain} is diagonal.
Strain applied along a general direction can lead to nonzero components of both $\alpha_x$ and $\alpha_y$ when the strain tensor is non-diagonal (see \secref{angle}). 


\section{Results}\label{sec:results}
\subsection{Electrical tuning}\label{sec:susc}
 
\begin{figure}
\centeredgraphics{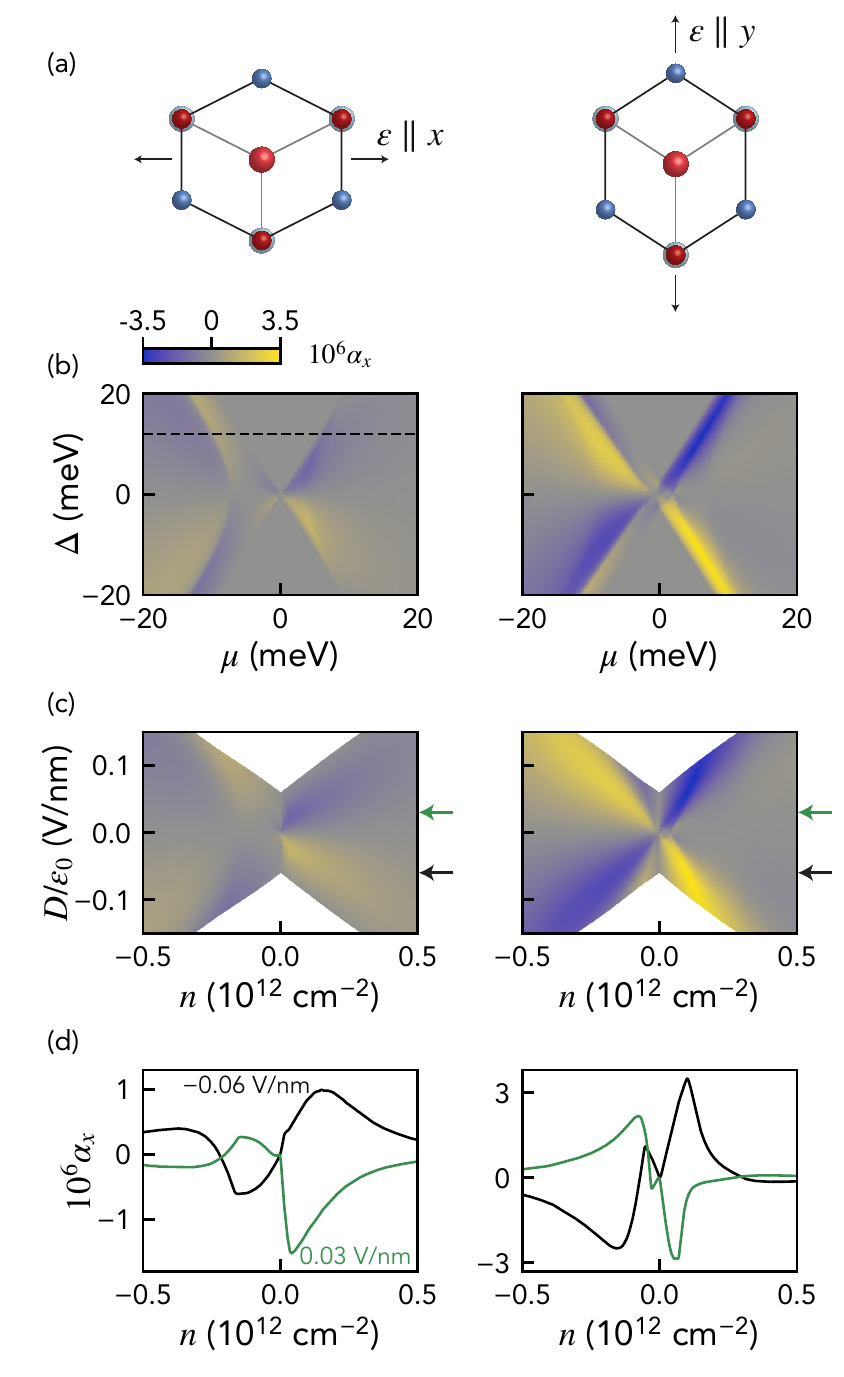}
\caption[Electrical tuning of magnetoelectric susceptibility]{
Magnetoelectric susceptibility $\alpha_x$ for strain of magnitude $\varepsilon=0.01$ aligned along the $x$ axis (left panels) and $y$ axis (right panels).
(a) Schematic of each strain state (not to scale).
(b) Maps plotted against model parameters $\Delta$ and $\mu$.
(c) The same maps as in (b) transformed onto axes of derived parameters $D$ and $n$.
The white triangular regions correspond to values of $D$ and $n$ not covered by the range of $\Delta$ and $\mu$ considered in (b).
(d) Line profiles for fixed $D/\varepsilon_0$ at positions indicated by the arrows in (c).
}
\label{fig:susc}
\end{figure}

We focus on the two strain orientations shown in \figref[a]{susc}, with strain either along the $x$ or $y$ direction for which $\alpha_y=0$.
For each strain orientation, \figref[b]{susc} shows $\alpha_x$ as a function of $\Delta$ and $\mu$. 
In an experiment, electrostatic gating directly adjusts the interlayer displacement field $D$ and carrier density $n$ rather than $\Delta$ and $\mu$ (see \appref{gates}). 
In \figref[c]{susc} we therefore show similar maps of $\alpha_x$ as a function of $D$ and $n$. 
The distortion of the features is due to the nonlinear relationships between $(\Delta , \mu)$ and $(D, n)$. 

The susceptibility $\alpha_x$ exhibits a rich dependence on these tuning parameters. 
The maps in \figref[a,c]{susc} are antisymmetric upon reversal of $\Delta$ or $D$, show non-monotonic dependences on the parameters, and have no symmetry between $\mu > 0$ and $\mu<0$ reflecting the lack of electron-hole symmetry of the Hamiltonian.
In both strain configurations, $\alpha_x$ reaches a broad maximum centered at larger $|n|$ for larger $|D|$, as highlighted by the line profiles in \figref{susc}(d).
Notably, the sign of $\alpha_x$ is fairly uniform in each of the four quadrants of the map, but also exhibits a sharp reversal in the valence band ($\mu<0$ or $n<0$).
This suggests that the orientation of $M_z$ can be reversed upon changing the carrier type, reversing the direction of the displacement field, or applying even smaller perturbations to either $n$ or $D$ in some regimes.

\begin{figure}
\centeredgraphics{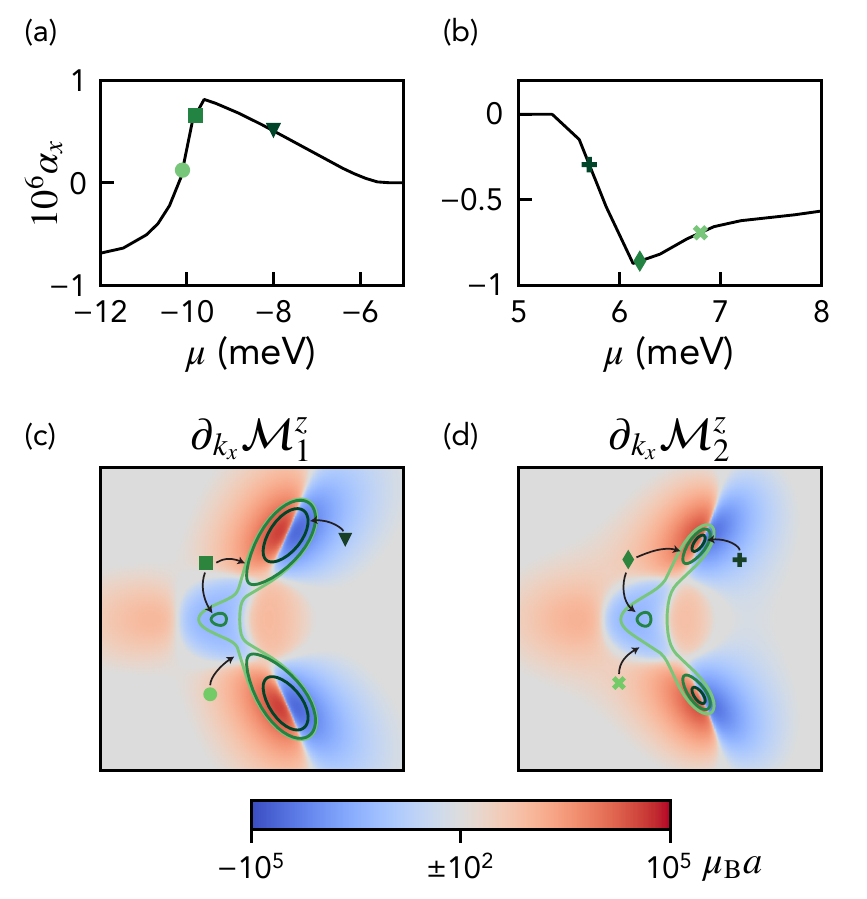}
\caption[Topological Lifshitz transitions of the Fermi surface]{
(a) Valence band and (b) conduction band line profiles of $\alpha_x$ from \figref{susc}(b) at $\Delta=12$~meV.
(c) $\partial_{k_x}\M_1^z$ at $\mu=-10.1$~meV and (d) $\partial_{k_x}\M_2^z$ at $6.8$~meV, with Fermi surface contours at the values of $\mu$ marked in (a)-(b).
From left to right: circle, $-10.1$~meV; square, $-9.8$~meV; triangle, $-8.0$~meV; ``+'', $5.7$~meV; diamond, $6.2$~meV; ``$\times$'', $6.8$~meV.
}
\label{fig:topo}
\end{figure}

The distinctive features in $\alpha_x$ coincide with changes in the topology of the Fermi surface. \Figref[a,b]{topo} shows line profiles from the left panel of \figref[b]{susc} at fixed $\Delta$ (dashed line), and \figref[c,d]{topo} shows the Fermi surfaces for a few values of $\mu$ superimposed on the typical momentum-space distribution of $\partial_{k_x} \MM^z$.
For small $|\mu|$ near the band edges, the Fermi surface first consists of two pockets (darkest contours) approximately centered around the hotspots of $\M$.
Sweeping to larger $|\mu|$, a central third pocket appears, and the three pockets eventually merge into a single continuous Fermi surface (lightest contour).
The appearance of the third pocket coincides with a cusp in $\alpha_x$, and the merging of the pockets coincides with an inflection point.
In the valence band, $\alpha_x$ changes sign approximately at this inflection point.
Our observations are consistent with Ref.~\cite{berrydipole}, which predicts a sign change in the Berry curvature dipole as a function of the carrier density in sBLG.


\begin{figure*}
\centeredgraphics{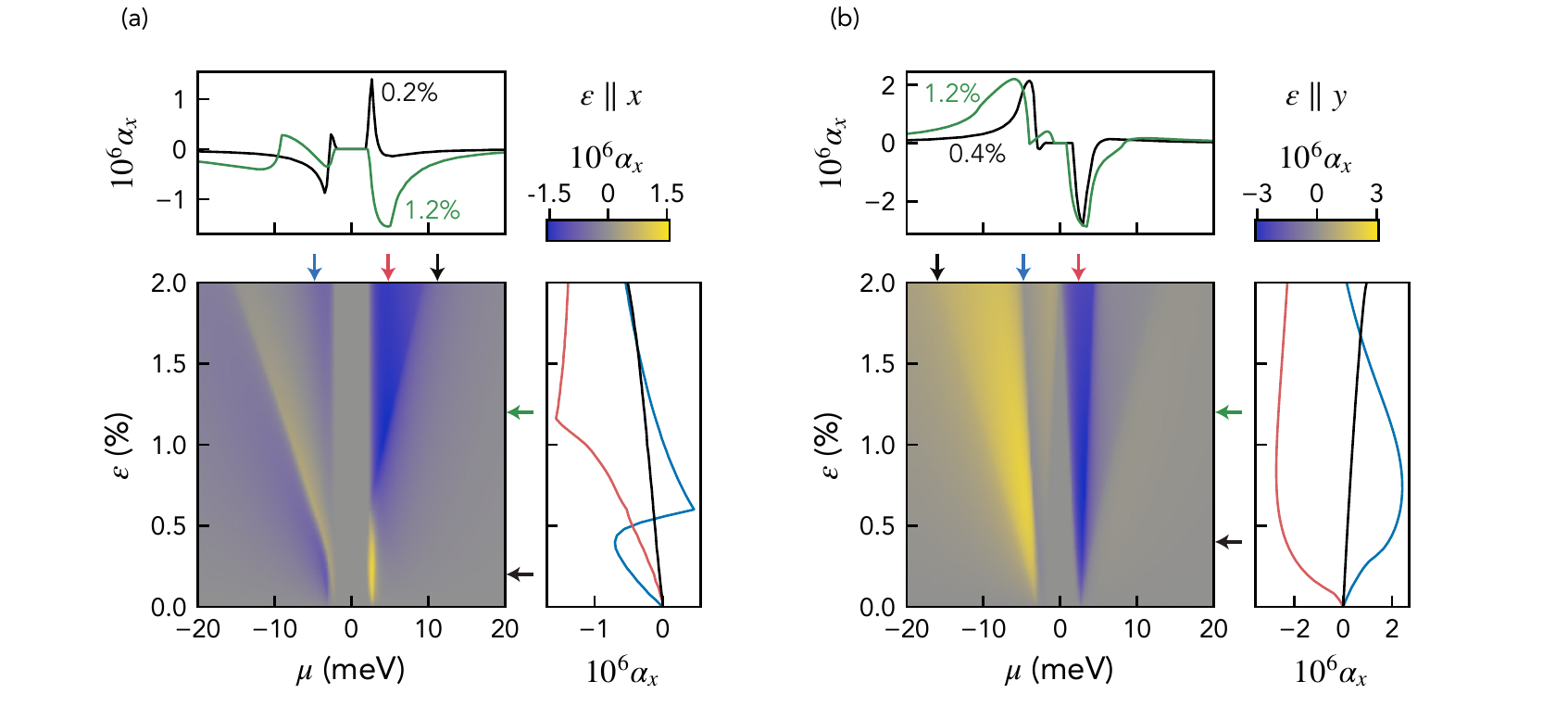}
\caption[Strain tuning of magnetoelectric susceptibility]{
Susceptibility $\alpha_x$ versus strain magnitude $\varepsilon$ and chemical potential $\mu$ for fixed $\Delta=5$~meV, with strain along either (a) the $x$ direction or (b) the $y$ direction. 
The arrows above each main panel indicate the following values of $\mu$ used for the line profiles in the corresponding right panel (in meV, from left to right): (a) $-4.8$, $4.8$, $11.2$; (b) $-16.0$, $-4.8$, $2.4$.
The arrows to the right of each main panel indicate the following values of $\varepsilon$) used for the line profiles in the corresponding upper panel (from top to bottom): (a) \SI{1.2}{\percent}, \SI{0.2}{\percent}; (b) \SI{1.2}{\percent}, \SI{0.4}{\percent}.
}
\label{fig:strain}
\end{figure*}

\subsection{Strain magnitude tuning}\label{sec:strain}

The strain amplitude $\varepsilon$ alters the band structure and $\MM$ distribution nontrivially.
In \figref{strain}, we show $\alpha_x$ as a function of $\mu$ and $\varepsilon$ for a fixed value of $\Delta$, again for strain applied along the $x$ and $y$ directions.
Similar to the previous section, we observe sharp features and non-monotonic dependences on $\varepsilon$ that are associated with changes in the Fermi surface topology. 
These changes generally occur at larger values of $|\mu|$ with increasing strain, because one of the three mini-valleys is shifted to higher energies. 
Therefore the corresponding Fermi pocket emerges at larger $|\mu|$.
At sufficiently large strain along the $y$ direction, the indirect band gap in sBLG closes. 
This gives rise to finite values of $\alpha_x$ at any value of $\mu$ for large strains as visible in \figref[b]{strain}.

The maximum value of $|\alpha_x|$ does not change significantly with $\varepsilon$. 
However, for large $|\mu|$ at which the Fermi surface consists of a single pocket, $\alpha_x$ is approximately proportional to $\varepsilon$ (see black curves in \figref{strain}).
This monotonic dependence of $\alpha_x$ on $\varepsilon$ is consistent with the magnetoelectric effect previously reported in strained monolayer MoS$_2$, which has a larger and approximately circular Fermi surface~\cite{mos2pull, mos2bend}.

\subsection{Strain angle tuning}\label{sec:angle}
 
\begin{figure}
\centeredgraphics{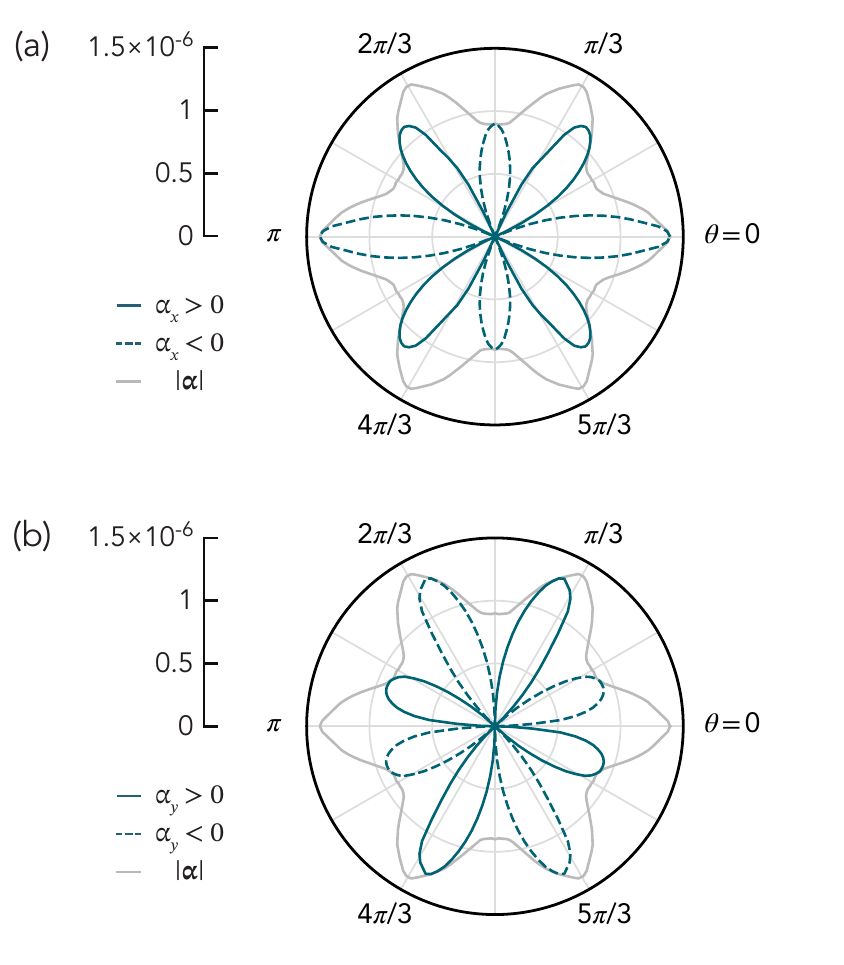}
\caption{
Dependence of (a) $\alpha_x$, (b) $\alpha_y$ (dark curves), and $|\aalpha|$ (light grey curve in both panels) on the angle $\theta$ between the principal strain axis and the $x$ axis.
The radial coordinate represents $|\aalpha|$, $\alpha_x$, or $\alpha_y$ up to a maximum of $1.5\times10^{-6}$.
The solid (dashed) part of the blue curves indicate where $\alpha_x$ or $\alpha_y$ is positive (negative).
In both panels, the model parameters are $\Delta=4$~meV, $\mu=4$~meV, and $\varepsilon=0.01$.
}
\label{fig:angle}
\end{figure}

Next, we show how $\aalpha$ depends on the orientation of the principal strain axis relative to the crystallographic axes.
\Figref{angle} plots the components of $\alpha_x$ and $\alpha_y$ as a function of the strain angle $\theta$ defined in \figref{tb}.
We also show the magnitude $|\aalpha| = \sqrt{\alpha_x^2+\alpha_y^2}$, which exhibits the six-fold symmetry of the unstrained lattice as expected.
The magnitude of the $x$ component $|\alpha_x|$ exhibits a local maximum whereas $\alpha_y=0$ for $\theta=0, \pi/2, ...$ at which the strain tensor is diagonal. This is due to our choice of coordinate system, in which $x$ and $y$ correspond to one of the zigzag and armchair directions of the lattice respectively. Generally, the shape and size of the lobes, in addition to the intermediate angles for which either component is zero, depend on the model parameters $\Delta$, $\mu$, and $\varepsilon$.

Applying an in-plane electric field $\EE$ generates an out-of-plane magnetization $M_z\propto \aalpha\cdot\EE$ (\eqref{ME}).
Therefore, the magnitude of $M_z$ is maximized with $\aalpha\parallel\EE$. 
We discuss below a relatively simple device geometry, with one pair of contacts that both clamps the sheet to apply strain and makes electrical contact to apply an in-plane electric field that drives a bias current (see \secref{experiment}).
In this geometry, the effect is maximum if a zigzag axis of the BLG crystal is aligned with the strain and current direction, which can be established during device fabrication.

\section{Discussion}\label{sec:discussion}

\subsection{Magnitude of the effect}\label{sec:magnitude}

Here we estimate the net magnetization that can be induced for realistic experimental parameters. 
For a device with channel width $W$, sheet resistance $\rho$ and a bias current $I$, \eqref{ME} can be rewritten as
\begin{equation}\label{eq:MoverI}
M_z = \frac{\tau\alpha_x\rho}{W\mu_0}I. 
\end{equation}
In our model, we use the linear relaxation-time approximation. 
This assumes that the shift $|\delta \kk| = e\tau|\EE|/\hbar$ of the Fermi surface does not exceed its momentum-space width (typically $0.01a^{-1}=\SI{7e7}{\per\meter}$, from \figref{topo}).
We also assume a scattering time $\tau\sim2$~ps for graphene-based systems near charge neutrality~\cite{relaxationtime1, relaxationtime2, relaxationtime3}. 
Together, this limits the maximum electric field strength $|\EE| < \SI{23000}{\volt\per\meter}$. 
We choose $|\EE|^{\text{max}}=\SI{e4}{\volt\per\meter}$ to be concrete. 
Further assuming $W\sim\SI{1}{\micron}$, $\rho\sim\SI{1}{\kilo\ohm}$, and a maximum $\alpha_x \sim\SI{3e-6}{}$ for \SI{1}{\percent} uniaxial strain (\figref{susc}), we arrive at $M_z \sim 0.005I$. 
This is expected to describe the system up to a maximum bias current $I^{\text{max}} = |\EE|^{\max} W/\rho \sim \SI{10}{\micro\ampere}$, corresponding to a maximum magnetization of $\SI{50}{\nano\ampere} = \SI{5400}{\magperum}$.

The magnetoelectric effect considered in this work has been theoretically predicted in strained monolayer NbSe$_2$~\cite{nbx2} and TBG~\cite{he} and experimentally observed in strained monolayer MoS$_2$~\cite{mos2pull, mos2bend}.
Here we briefly compare the magnitude of the effect for these different materials assuming \tsim0.5-\SI{1}{\percent} uniaxial tensile strain (see \appref{magcomparison} for details). 
We find $M_z/I \sim \SI{5e-6}{}$ in MoS$_2$, $M_z/I \sim \SI{e-3}{}$ in NbSe$_2$, and $M_z/I \sim \SI{4e-4}{}$ in TBG.
In sBLG, we predict a potentially larger effect with magnetization as high as $M_z/I \sim \SI{5e-3}{}$ possible in some regimes of the tuning parameters.
This is a result of the large magnitude and asymmetric redistribution of the orbital magnetic moment and Berry curvature in sBLG.

\subsection{Proposed experimental observation}\label{sec:experiment}

Studying the magnetoelectric effect in sBLG experimentally requires two components: (1) a technique to strain dual-gated BLG devices with electrical contacts and (2) a technique to detect the resultant magnetization.
Recently, several experimental approaches have been reported to continuously and reversibly strain devices based on two-dimensional materials while maintaining well-performing electrical contacts~\cite{lujun, lujun2, lujun3, mos2bend}.
In MoS$_2$ ~\cite{mos2bend, mos2pull}, the magnetization was probed previously using magneto-optic imaging, but due to the small bandgap in BLG this is challenging to apply here.
We therefore consider scanning magnetometry techniques that detect the stray magnetic field above the surface of a device such as scanning SQUID, Hall probe, or nitrogen-vacancy center microscopy~\cite{sot, kirtleyreview, nvreview, thiel, sotmoire}.

\begin{figure}
\centeredgraphics{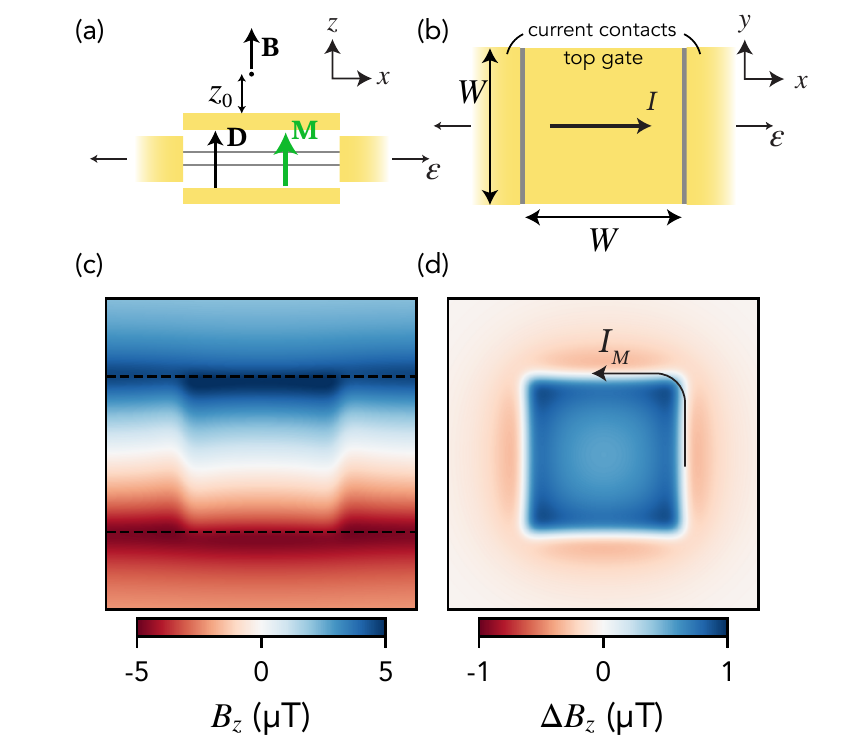}
\caption{
(a) Side view and (b) top view schematics of a \SI{1x1}{\micron} square sBLG device with orbital magnetization $\mathbf{M}=M_z\hat{z}$.
The metal electrodes simultaneously clamp the BLG sheet and bias the device with a current $I$ in the $+x$ direction. 
Applying strain to the substrate along the $x$ direction results in corresponding strain in the BLG sheet aligned with the applied current.
The top and bottom metal gates tune the carrier density $n$ and displacement field $D$.
(c) Total out-of-plane stray magnetic field $B_z$ at a height $z_0=100$~nm above the surface of the device for $M_z/I=0.05$ and $I=10$~\si{\micro\ampere}.
(d) Difference in magnetic field $\Delta B_z = \left[B_z (D) - B_z(-D)\right]/2$ between images at opposite $D$.
}
\label{fig:detect}
\end{figure}

\Figref[a,b]{detect} shows a schematic of a $W\times W$ square sBLG device that is strained and electrically biased with the same pair of metal contacts.
Applying voltage to the top and bottom metal gates tunes the electric displacement field $D$ and carrier density $n$.
The total magnetic field above the device is a combination of the Oersted field due to the bias current $B_{\text{bias}}(\mathbf{r})$ and the magnetic field produced by the orbital magnetization $B_{\text{sBLG}}(\mathbf{r})$:
$$
B_z(\mathbf{r}) = B_{\text{bias}}(\mathbf{r}) + B_{\text{sBLG}}(\mathbf{r}).
$$
We model $B_{\text{bias}}(\mathbf{r})$ using an infinitely long, width-$W$ wire with current flowing in the $+x$ direction.
Because the induced magnetization is out-of-plane, $B_{\text{sBLG}}(\mathbf{r})$ can be modeled using the Oersted field from an effective current $I_M$ with magnitude $M_z$ flowing at the boundary of the sBLG square.

\Figref[c]{detect} shows the $z$ component of the stray magnetic field at height $z_0=100$~nm above the surface of a sBLG device.
$B_{\text{bias}}(\mathbf{r})$ dominates the image, with a distortion from $B_{\text{sBLG}}(\mathbf{r})$.
The contrast between the two sources of field is controlled by the ratio $M_z/I$. 
For illustration, we use a value of $M_z/I$ ten times larger than estimated above.
To clearly reveal the magnetization, we take advantage of the symmetry $M_z(D)=-M_z(-D)$.
Reversing the sign of $D$ reverses the sign of $B_{\text{sBLG}}(\mathbf{r})$ without changing the Oersted field from the bias current. 
The difference of two images corresponding to opposite values of $D$ therefore shows the stray field from the magnetization alone (\figref[d]{detect}). 
From \figref[d]{detect}, we see that the magnetic field is on the order of tens to hundreds of nanotesla (accounting for the factor of 10 by which the effect is exaggerated). 
Detecting this magnetic field strength is within the capabilities of scanning magnetometry techniques, with typical best magnetic field sensitivity down to the nanotesla scale~\cite{kirtleyreview}.

\section{Conclusion}\label{sec:conclusion}
In summary, we develop a tight-binding model for sBLG that predicts an orbital magnetization on the order of up to \SI{5000}{\magperum} under a \SI{1}{\percent} uniaxial strain and \SI{10}{\micro\ampere} bias current.
The model includes all relevant nearest and next-nearest neighbor coupling terms and is compatible with an arbitrary direction of the applied strain.
We show that the effect is a source of an electrically controlled out-of-plane magnetization that is uniform throughout the sBLG layer and can be experimentally detected using scanning magnetometry. 
One opportunity to explore this effect is in the context of 2D spintronic devices~\cite{2dspintronicreview}. 
Here, an interesting possibility is to combine sBLG with a magnetic layer and explore switching of the layer by current-induced magnetic torques from sBLG. 
Finally, the orbital magnetoelectric effect discussed here offers a direct window into Berry curvature effects in BLG, which are of both fundamental and applied interest.

\begin{acknowledgments}
The authors thank Kin Fai Mak and Erich Mueller for insightful discussions.
This work was primarily supported by the Cornell Center for Materials Research with funding from the NSF MRSEC program (DMR-1719875).
\end{acknowledgments}

\appendix
\section{Effective two-band Hamiltonian}\label{app:twoband}
For monolayer graphene, it is an established result that the Berry curvature and orbital magnetic moment inversely depends on the band gap~\cite{valleycontrasting}. 
Here we briefly derive the same result for a simple model of BLG in the absence of strain to show that the same general dependence is expected in BLG. 
A low-energy two-band effective Hamiltonian for electronic states at location $\mathbf{q}=(q_x,q_y)$ in momentum space relative to the K or K$'$ point is~\cite{mccann}:
$$
H = -\frac{\hbar^2}{2m}\left(\begin{array}{cc} 
0 & (q_x - iq_y)^2 \\ 
(q_x + iq_y)^2 & 0
\end{array}\right)
+
\frac{\Delta}{2}\left(\begin{array}{cc} 
1 & 0 \\
0 & -1
\end{array}\right),
$$
where $$m=\frac{\gamma_1}{2v^2}$$ is an effective mass, with $$v=\frac{\sqrt{3}}{2}\frac{a\gamma_0}{\hbar}.$$
This Hamiltonian only includes the hopping parameters $\gamma_0$ and $\gamma_1$ (\tabref{hopping}) and therefore lacks trigonal warping and is electron-hole symmetric.
Diagonalizing this Hamiltonian leads to symmetric parabolic energy bands with a mass gap $\Delta$:
$$
E_n(\kk) = \pm\sqrt{\left(\frac{\hbar^2|\kk|^2}{2m}\right)^2 + \left(\frac{\Delta}{2}\right)^2}.
$$
For any two-band model with particle-hole symmetry~\cite{valleycontrasting},
$$
\m(\kk)\equiv\frac{e}{\hbar}E_n(\kk)\berry(\kk)
$$
and \eqref{Mn} becomes
$$
\M(\kk) = \frac{e}{\hbar}\mu\berry(\kk),
$$
where $\mu$ is the chemical potential.
Using the expression for the Berry curvature reported previously~\cite{twobandmoment, twobandmoment2, valleycurrentBLG}, we obtain 
$$
\M(\kk) = \mp\frac{e\hbar}{2m}\frac{\mu\Delta\sqrt{E_n(\kk)^2-\left(\frac\Delta2\right)^2}}{E_n(\kk)^3},
$$
where the upper (lower) sign corresponds to the K (K$'$) valley.
Under this model, $\M$ is zero at the valley center and is distributed in a ring at finite momentum
$$
|\kk^*| = \sqrt{\frac{m}{\sqrt2}}\frac{\Delta}{\hbar} 
$$
surrounding the valley center.
At this momentum, $\M$ reaches a maximum value of
$$
\M^{\text{max}} = \mp \frac{8}{3\sqrt{3}}\frac{e\hbar}{2m}\frac{\mu}{\Delta}
$$
which depends on the ratio of the chemical potential to interlayer asymmetry ($\mu/\Delta$).
This expression implies that BLG systems with smaller $\Delta$ will exhibit stronger valley effects.

\section{Electrostatic tuning of model parameters}\label{app:gates}
In the calculations discussed above, we fix the interlayer asymmetry $\Delta$ and chemical potential $\mu$.
Experimentally, electrostatic tuning controls instead the interlayer electric displacement field $D$ and carrier density $n$~\cite{BLGgap, valleycurrentBLG, valleycurrentBLG2}. 
The total carrier density $n=n_1+n_2$ is the sum of carrier density $n_i$ on each layer, with~\cite{mccann}:
\begin{equation}\label{eq:ni}
n_i = 4 \sum_n \int \frac{\dd\kk}{(2\pi)^2} f^0_n(\kk)\left[\left|\psi_n^{Ai}\right|^2 + \left|\psi_n^{Bi}\right|^2\right].
\end{equation}
The displacement field $D$ is related to $\Delta$ accounting for screening from charge on each layer~\cite{interlayerasymmetry, mccann}:
\begin{equation}\label{eq:D}
    \Delta = \frac{ed}{\epsilon_0}\left[D + (n_1 - n_2)e\right],
\end{equation}
where $d=0.34$~nm is the interlayer spacing.
We use \eqref{ni} and \eqref{D} along with the results of our model for chosen $\Delta$ and $\mu$ to obtain \figref[c]{susc} from \figref[b]{susc}.

Considering the device structure in \figref[a-b]{detect}, the displacement field $D$ is a difference between gate voltages $V_1$ and $V_2$, while the carrier density $n$ is essentially a sum of the gate voltages.
Accounting for different dielectric constants ($\epsilon_1$, $\epsilon_2$) and dielectric layer thicknesses ($d_1$, $d_2$)~\cite{BLGgap}: 
$$
D = \frac{\epsilon_0 \epsilon_1 V_1}{d_1} - \frac{\epsilon_0 \epsilon_2 V_2}{d_2}
$$
$$
n e = (n_1 +n_2)e = \frac{\epsilon_0 \epsilon_1 V_1}{d_1} + \frac{\epsilon_0 \epsilon_2 V_2}{d_2}.
$$
Solving for the gate voltages directly, 
$$
\frac{\epsilon_0 \epsilon_1 V_1}{d_1} = \frac12\frac{\epsilon_0 \Delta}{de} + n_2 e
$$
$$
\frac{\epsilon_0 \epsilon_2 V_2}{d_2} =  -\frac12\frac{\epsilon_0 \Delta}{de} + n_1 e.
$$
These expressions are useful for comparing the expected magnitude of the magnetoelectric susceptibility calculated here to that obtained experimentally under equivalent conditions.
The inverse problem (determining $\Delta$, $\mu$ from gate voltages $V_1$, $V_2$) is less approachable, since the calculation of $n_1$ and $n_2$ requires integration over a Fermi surface with nontrivial geometry and topology (see \eqref{ni}).

\section{Magnitude of orbital magnetization in other systems}\label{app:magcomparison}
Here we compare the orbital magnetization predicted for sBLG to a few other 2D materials. 
The magnetoelectric effects described below are expected to be linear in both strain magnitude $\varepsilon$ and bias current $I$, so we calculate the normalized magnetization $M_z/(I\varepsilon)$.
\subsection{MoS$_2$}
In Ref.~\cite{mos2pull}, the combination of Kerr rotation microscopy of strained single-layer MoS$_2$ and a tight-binding model is used to estimate $M_z\sim\SI{4e-11}{\ampere}$ with current density $J\sim\SI[per-mode=symbol]{10}{\ampere\per\meter}$ for a $W\sim\SI{8}{\micro\meter}$ device under $\varepsilon\sim\SI{0.5}{\percent}$ strain~\cite{mos2pull}.
This is equivalent to a bias current of $I\sim\SI{80}{\micro\ampere}$ and normalized magnetization $M_z/(I\varepsilon)\sim\SI[per-mode=symbol]{e-6}{\ampere\per\ampere\per\percent}$.
Ref.~\cite{mos2bend} reports a maximum estimated volume magnetization per unit current density $(M_z/t)/(I/W)\sim0.1$ for a $W\sim\SI{12}{\micro\meter}$ device with thickness $t=0.67$~nm under $\varepsilon\sim\SI{1}{\percent}$ strain.
This corresponds to a normalized area magnetization $M_z/(I\varepsilon) = 0.1(t/W)/(\SI{1}{\percent}) = \SI[per-mode=symbol]{5e-6}{\ampere\per\ampere\per\percent}$. \\

\subsection{NbSe$_2$}
A tight-binding model for strained monolayer NbSe$_2$~\cite{nbx2} predicts $M_z\sim\SI{e4}{\magperum}\sim\SI{e-7}{\ampere}$ for $\varepsilon=\SI{5}{\percent}$ and $\mathcal{E}=\SI{e4}{\volt\per\meter}$.
Assuming a resistivity $\rho\sim\SI{1}{\kilo\ohm}$ and device length $L\sim\SI{10}{\micro\meter}$, this corresponds to an approximate bias current $I\sim\SI{100}{\micro\ampere}$ and normalized magnetization $M_z/(I\varepsilon)\sim \SI[per-mode=symbol]{2e-4}{\ampere\per\ampere\per\percent}$.

\subsection{Twisted bilayer graphene}
Ref.~\cite{he} predicts a magnetoelectric effect in strained TBG with a relative rotation of \SI{1.2}{\degree} between the layers and with the symmetry between the layers further broken by a hexagonal boron nitride substrate~\cite{he}.
For \SI{0.1}{\percent} uniform uniaxial strain and $\mathcal{E}=\SI{e4}{\volt\per\meter}$, the estimated magnitude of $M_z$ is \tsim$\SI{2e4}{\magperum}\sim\SI{2e7}{\ampere}$.
Using $\rho\sim\SI{1}{\kilo\ohm}$ and $L\sim\SI{10}{\micron}$, the normalized magnetization is $M_z/(I\varepsilon)\sim\SI[per-mode=symbol]{4e-4}{\ampere\per\ampere\per\percent}$.

\bibliographystyle{apsrev4-2}
\bibliography{main}

\begin{thebibliography}{62}%
\makeatletter
\providecommand \@ifxundefined [1]{%
 \@ifx{#1\undefined}
}%
\providecommand \@ifnum [1]{%
 \ifnum #1\expandafter \@firstoftwo
 \else \expandafter \@secondoftwo
 \fi
}%
\providecommand \@ifx [1]{%
 \ifx #1\expandafter \@firstoftwo
 \else \expandafter \@secondoftwo
 \fi
}%
\providecommand \natexlab [1]{#1}%
\providecommand \enquote  [1]{``#1''}%
\providecommand \bibnamefont  [1]{#1}%
\providecommand \bibfnamefont [1]{#1}%
\providecommand \citenamefont [1]{#1}%
\providecommand \href@noop [0]{\@secondoftwo}%
\providecommand \href [0]{\begingroup \@sanitize@url \@href}%
\providecommand \@href[1]{\@@startlink{#1}\@@href}%
\providecommand \@@href[1]{\endgroup#1\@@endlink}%
\providecommand \@sanitize@url [0]{\catcode `\\12\catcode `\$12\catcode
  `\&12\catcode `\#12\catcode `\^12\catcode `\_12\catcode `\%12\relax}%
\providecommand \@@startlink[1]{}%
\providecommand \@@endlink[0]{}%
\providecommand \url  [0]{\begingroup\@sanitize@url \@url }%
\providecommand \@url [1]{\endgroup\@href {#1}{\urlprefix }}%
\providecommand \urlprefix  [0]{URL }%
\providecommand \Eprint [0]{\href }%
\providecommand \doibase [0]{https://doi.org/}%
\providecommand \selectlanguage [0]{\@gobble}%
\providecommand \bibinfo  [0]{\@secondoftwo}%
\providecommand \bibfield  [0]{\@secondoftwo}%
\providecommand \translation [1]{[#1]}%
\providecommand \BibitemOpen [0]{}%
\providecommand \bibitemStop [0]{}%
\providecommand \bibitemNoStop [0]{.\EOS\space}%
\providecommand \EOS [0]{\spacefactor3000\relax}%
\providecommand \BibitemShut  [1]{\csname bibitem#1\endcsname}%
\let\auto@bib@innerbib\@empty
\bibitem [{\citenamefont {{Castro Neto}}\ \emph {et~al.}(2009)\citenamefont
  {{Castro Neto}}, \citenamefont {Guinea}, \citenamefont {Peres}, \citenamefont
  {Novoselov},\ and\ \citenamefont {Geim}}]{theoryreview}%
  \BibitemOpen
  \bibfield  {author} {\bibinfo {author} {\bibfnamefont {A.~H.}\ \bibnamefont
  {{Castro Neto}}}, \bibinfo {author} {\bibfnamefont {F.}~\bibnamefont
  {Guinea}}, \bibinfo {author} {\bibfnamefont {N.~M.~R.}\ \bibnamefont
  {Peres}}, \bibinfo {author} {\bibfnamefont {K.~S.}\ \bibnamefont
  {Novoselov}},\ and\ \bibinfo {author} {\bibfnamefont {A.~K.}\ \bibnamefont
  {Geim}},\ }\href {https://doi.org/10.1103/RevModPhys.81.109} {\bibfield
  {journal} {\bibinfo  {journal} {Rev. Mod. Phys.}\ }\textbf {\bibinfo {volume}
  {81}},\ \bibinfo {pages} {109} (\bibinfo {year} {2009})}\BibitemShut
  {NoStop}%
\bibitem [{\citenamefont {Xiao}\ \emph {et~al.}(2010)\citenamefont {Xiao},
  \citenamefont {Chang},\ and\ \citenamefont {Niu}}]{berryreview}%
  \BibitemOpen
  \bibfield  {author} {\bibinfo {author} {\bibfnamefont {D.}~\bibnamefont
  {Xiao}}, \bibinfo {author} {\bibfnamefont {M.-C.}\ \bibnamefont {Chang}},\
  and\ \bibinfo {author} {\bibfnamefont {Q.}~\bibnamefont {Niu}},\ }\href
  {https://doi.org/10.1103/RevModPhys.82.1959} {\bibfield  {journal} {\bibinfo
  {journal} {Rev. Mod. Phys.}\ }\textbf {\bibinfo {volume} {82}},\ \bibinfo
  {pages} {1959} (\bibinfo {year} {2010})}\BibitemShut {NoStop}%
\bibitem [{\citenamefont {Xiao}\ \emph {et~al.}(2007)\citenamefont {Xiao},
  \citenamefont {Yao},\ and\ \citenamefont {Niu}}]{valleycontrasting}%
  \BibitemOpen
  \bibfield  {author} {\bibinfo {author} {\bibfnamefont {D.}~\bibnamefont
  {Xiao}}, \bibinfo {author} {\bibfnamefont {W.}~\bibnamefont {Yao}},\ and\
  \bibinfo {author} {\bibfnamefont {Q.}~\bibnamefont {Niu}},\ }\href
  {https://doi.org/10.1103/PhysRevLett.99.236809} {\bibfield  {journal}
  {\bibinfo  {journal} {Phys. Rev. Lett.}\ }\textbf {\bibinfo {volume} {99}},\
  \bibinfo {pages} {236809} (\bibinfo {year} {2007})}\BibitemShut {NoStop}%
\bibitem [{\citenamefont {Ahn}(2020)}]{2dspintronicreview}%
  \BibitemOpen
  \bibfield  {author} {\bibinfo {author} {\bibfnamefont {E.~C.}\ \bibnamefont
  {Ahn}},\ }\href {https://doi.org/10.1038/s41699-020-0152-0} {\bibfield
  {journal} {\bibinfo  {journal} {npj 2D Mater. Appl.}\ }\textbf {\bibinfo
  {volume} {4}},\ \bibinfo {pages} {17} (\bibinfo {year} {2020})}\BibitemShut
  {NoStop}%
\bibitem [{\citenamefont {Mak}\ \emph {et~al.}(2018)\citenamefont {Mak},
  \citenamefont {Xiao},\ and\ \citenamefont {Shan}}]{faijiereview}%
  \BibitemOpen
  \bibfield  {author} {\bibinfo {author} {\bibfnamefont {K.~F.}\ \bibnamefont
  {Mak}}, \bibinfo {author} {\bibfnamefont {D.}~\bibnamefont {Xiao}},\ and\
  \bibinfo {author} {\bibfnamefont {J.}~\bibnamefont {Shan}},\ }\href
  {https://doi.org/10.1038/s41566-018-0204-6} {\bibfield  {journal} {\bibinfo
  {journal} {Nat. Photonics}\ }\textbf {\bibinfo {volume} {12}},\ \bibinfo
  {pages} {451} (\bibinfo {year} {2018})}\BibitemShut {NoStop}%
\bibitem [{\citenamefont {Mak}\ \emph {et~al.}(2014)\citenamefont {Mak},
  \citenamefont {McGill}, \citenamefont {Park},\ and\ \citenamefont
  {McEuen}}]{valleyhall}%
  \BibitemOpen
  \bibfield  {author} {\bibinfo {author} {\bibfnamefont {K.~F.}\ \bibnamefont
  {Mak}}, \bibinfo {author} {\bibfnamefont {K.~L.}\ \bibnamefont {McGill}},
  \bibinfo {author} {\bibfnamefont {J.}~\bibnamefont {Park}},\ and\ \bibinfo
  {author} {\bibfnamefont {P.~L.}\ \bibnamefont {McEuen}},\ }\href
  {https://doi.org/10.1126/science.1250140} {\bibfield  {journal} {\bibinfo
  {journal} {Science}\ }\textbf {\bibinfo {volume} {344}},\ \bibinfo {pages}
  {1489} (\bibinfo {year} {2014})}\BibitemShut {NoStop}%
\bibitem [{\citenamefont {Lee}\ \emph {et~al.}(2016)\citenamefont {Lee},
  \citenamefont {Mak},\ and\ \citenamefont {Shan}}]{valleyhalljieun}%
  \BibitemOpen
  \bibfield  {author} {\bibinfo {author} {\bibfnamefont {J.}~\bibnamefont
  {Lee}}, \bibinfo {author} {\bibfnamefont {K.~F.}\ \bibnamefont {Mak}},\ and\
  \bibinfo {author} {\bibfnamefont {J.}~\bibnamefont {Shan}},\ }\href
  {https://doi.org/10.1038/nnano.2015.337} {\bibfield  {journal} {\bibinfo
  {journal} {Nat. Nanotechnol.}\ }\textbf {\bibinfo {volume} {11}},\ \bibinfo
  {pages} {421} (\bibinfo {year} {2016})}\BibitemShut {NoStop}%
\bibitem [{\citenamefont {Li}\ \emph {et~al.}(2020{\natexlab{a}})\citenamefont
  {Li}, \citenamefont {Jiang}, \citenamefont {Wang}, \citenamefont {Watanabe},
  \citenamefont {Taniguchi}, \citenamefont {Shan},\ and\ \citenamefont
  {Mak}}]{lizhong}%
  \BibitemOpen
  \bibfield  {author} {\bibinfo {author} {\bibfnamefont {L.}~\bibnamefont
  {Li}}, \bibinfo {author} {\bibfnamefont {S.}~\bibnamefont {Jiang}}, \bibinfo
  {author} {\bibfnamefont {Z.}~\bibnamefont {Wang}}, \bibinfo {author}
  {\bibfnamefont {K.}~\bibnamefont {Watanabe}}, \bibinfo {author}
  {\bibfnamefont {T.}~\bibnamefont {Taniguchi}}, \bibinfo {author}
  {\bibfnamefont {J.}~\bibnamefont {Shan}},\ and\ \bibinfo {author}
  {\bibfnamefont {K.~F.}\ \bibnamefont {Mak}},\ }\href
  {https://doi.org/10.1103/PhysRevMaterials.4.104005} {\bibfield  {journal}
  {\bibinfo  {journal} {Phys. Rev. Mater.}\ }\textbf {\bibinfo {volume} {4}},\
  \bibinfo {pages} {104005} (\bibinfo {year} {2020}{\natexlab{a}})}\BibitemShut
  {NoStop}%
\bibitem [{\citenamefont {Lee}\ \emph {et~al.}(2017)\citenamefont {Lee},
  \citenamefont {Wang}, \citenamefont {Xie}, \citenamefont {Mak},\ and\
  \citenamefont {Shan}}]{mos2pull}%
  \BibitemOpen
  \bibfield  {author} {\bibinfo {author} {\bibfnamefont {J.}~\bibnamefont
  {Lee}}, \bibinfo {author} {\bibfnamefont {Z.}~\bibnamefont {Wang}}, \bibinfo
  {author} {\bibfnamefont {H.}~\bibnamefont {Xie}}, \bibinfo {author}
  {\bibfnamefont {K.~F.}\ \bibnamefont {Mak}},\ and\ \bibinfo {author}
  {\bibfnamefont {J.}~\bibnamefont {Shan}},\ }\href
  {https://doi.org/10.1038/nmat4931} {\bibfield  {journal} {\bibinfo  {journal}
  {Nat. Mater.}\ }\textbf {\bibinfo {volume} {16}},\ \bibinfo {pages} {887}
  (\bibinfo {year} {2017})}\BibitemShut {NoStop}%
\bibitem [{\citenamefont {Son}\ \emph {et~al.}(2019)\citenamefont {Son},
  \citenamefont {Kim}, \citenamefont {Ahn}, \citenamefont {Lee},\ and\
  \citenamefont {Lee}}]{mos2bend}%
  \BibitemOpen
  \bibfield  {author} {\bibinfo {author} {\bibfnamefont {J.}~\bibnamefont
  {Son}}, \bibinfo {author} {\bibfnamefont {K.-H.}\ \bibnamefont {Kim}},
  \bibinfo {author} {\bibfnamefont {Y.~H.}\ \bibnamefont {Ahn}}, \bibinfo
  {author} {\bibfnamefont {H.-W.}\ \bibnamefont {Lee}},\ and\ \bibinfo {author}
  {\bibfnamefont {J.}~\bibnamefont {Lee}},\ }\href
  {https://doi.org/10.1103/PhysRevLett.123.036806} {\bibfield  {journal}
  {\bibinfo  {journal} {Phys. Rev. Lett.}\ }\textbf {\bibinfo {volume} {123}},\
  \bibinfo {pages} {036806} (\bibinfo {year} {2019})}\BibitemShut {NoStop}%
\bibitem [{\citenamefont {Zhang}\ \emph {et~al.}(2009)\citenamefont {Zhang},
  \citenamefont {Tang}, \citenamefont {Girit}, \citenamefont {Hao},
  \citenamefont {Martin}, \citenamefont {Zettl}, \citenamefont {Crommie},
  \citenamefont {Shen},\ and\ \citenamefont {Wang}}]{BLGgap}%
  \BibitemOpen
  \bibfield  {author} {\bibinfo {author} {\bibfnamefont {Y.}~\bibnamefont
  {Zhang}}, \bibinfo {author} {\bibfnamefont {T.-T.}\ \bibnamefont {Tang}},
  \bibinfo {author} {\bibfnamefont {C.}~\bibnamefont {Girit}}, \bibinfo
  {author} {\bibfnamefont {Z.}~\bibnamefont {Hao}}, \bibinfo {author}
  {\bibfnamefont {M.~C.}\ \bibnamefont {Martin}}, \bibinfo {author}
  {\bibfnamefont {A.}~\bibnamefont {Zettl}}, \bibinfo {author} {\bibfnamefont
  {M.~F.}\ \bibnamefont {Crommie}}, \bibinfo {author} {\bibfnamefont {Y.~R.}\
  \bibnamefont {Shen}},\ and\ \bibinfo {author} {\bibfnamefont
  {F.}~\bibnamefont {Wang}},\ }\href {https://doi.org/10.1038/nature08105}
  {\bibfield  {journal} {\bibinfo  {journal} {Nature}\ }\textbf {\bibinfo
  {volume} {459}},\ \bibinfo {pages} {820} (\bibinfo {year}
  {2009})}\BibitemShut {NoStop}%
\bibitem [{\citenamefont {Shimazaki}\ \emph {et~al.}(2015)\citenamefont
  {Shimazaki}, \citenamefont {Yamamoto}, \citenamefont {Borzenets},
  \citenamefont {Watanabe}, \citenamefont {Taniguchi},\ and\ \citenamefont
  {Tarucha}}]{valleycurrentBLG}%
  \BibitemOpen
  \bibfield  {author} {\bibinfo {author} {\bibfnamefont {Y.}~\bibnamefont
  {Shimazaki}}, \bibinfo {author} {\bibfnamefont {M.}~\bibnamefont {Yamamoto}},
  \bibinfo {author} {\bibfnamefont {I.~V.}\ \bibnamefont {Borzenets}}, \bibinfo
  {author} {\bibfnamefont {K.}~\bibnamefont {Watanabe}}, \bibinfo {author}
  {\bibfnamefont {T.}~\bibnamefont {Taniguchi}},\ and\ \bibinfo {author}
  {\bibfnamefont {S.}~\bibnamefont {Tarucha}},\ }\href
  {https://doi.org/10.1038/nphys3551} {\bibfield  {journal} {\bibinfo
  {journal} {Nat. Phys.}\ }\textbf {\bibinfo {volume} {11}},\ \bibinfo {pages}
  {1032} (\bibinfo {year} {2015})}\BibitemShut {NoStop}%
\bibitem [{\citenamefont {Sui}\ \emph {et~al.}(2015)\citenamefont {Sui},
  \citenamefont {Chen}, \citenamefont {Ma}, \citenamefont {Shan}, \citenamefont
  {Tian}, \citenamefont {Watanabe}, \citenamefont {Taniguchi}, \citenamefont
  {Jin}, \citenamefont {Yao}, \citenamefont {Xiao},\ and\ \citenamefont
  {Zhang}}]{valleycurrentBLG2}%
  \BibitemOpen
  \bibfield  {author} {\bibinfo {author} {\bibfnamefont {M.}~\bibnamefont
  {Sui}}, \bibinfo {author} {\bibfnamefont {G.}~\bibnamefont {Chen}}, \bibinfo
  {author} {\bibfnamefont {L.}~\bibnamefont {Ma}}, \bibinfo {author}
  {\bibfnamefont {W.-Y.}\ \bibnamefont {Shan}}, \bibinfo {author}
  {\bibfnamefont {D.}~\bibnamefont {Tian}}, \bibinfo {author} {\bibfnamefont
  {K.}~\bibnamefont {Watanabe}}, \bibinfo {author} {\bibfnamefont
  {T.}~\bibnamefont {Taniguchi}}, \bibinfo {author} {\bibfnamefont
  {X.}~\bibnamefont {Jin}}, \bibinfo {author} {\bibfnamefont {W.}~\bibnamefont
  {Yao}}, \bibinfo {author} {\bibfnamefont {D.}~\bibnamefont {Xiao}},\ and\
  \bibinfo {author} {\bibfnamefont {Y.}~\bibnamefont {Zhang}},\ }\href
  {https://doi.org/10.1038/nphys3485} {\bibfield  {journal} {\bibinfo
  {journal} {Nat. Phys.}\ }\textbf {\bibinfo {volume} {11}},\ \bibinfo {pages}
  {1027} (\bibinfo {year} {2015})}\BibitemShut {NoStop}%
\bibitem [{\citenamefont {Yankowitz}\ \emph {et~al.}(2019)\citenamefont
  {Yankowitz}, \citenamefont {Ma}, \citenamefont {Jarillo-Herrero},\ and\
  \citenamefont {LeRoy}}]{vdwYankowitz}%
  \BibitemOpen
  \bibfield  {author} {\bibinfo {author} {\bibfnamefont {M.}~\bibnamefont
  {Yankowitz}}, \bibinfo {author} {\bibfnamefont {Q.}~\bibnamefont {Ma}},
  \bibinfo {author} {\bibfnamefont {P.}~\bibnamefont {Jarillo-Herrero}},\ and\
  \bibinfo {author} {\bibfnamefont {B.~J.}\ \bibnamefont {LeRoy}},\ }\href
  {https://doi.org/10.1038/s42254-018-0016-0} {\bibfield  {journal} {\bibinfo
  {journal} {Nat. Rev. Phys.}\ }\textbf {\bibinfo {volume} {1}},\ \bibinfo
  {pages} {112} (\bibinfo {year} {2019})}\BibitemShut {NoStop}%
\bibitem [{\citenamefont {Zhou}\ \emph {et~al.}(2019)\citenamefont {Zhou},
  \citenamefont {Taguchi}, \citenamefont {Kawaguchi}, \citenamefont {Tanaka},\
  and\ \citenamefont {Law}}]{spinorbitTMD}%
  \BibitemOpen
  \bibfield  {author} {\bibinfo {author} {\bibfnamefont {B.~T.}\ \bibnamefont
  {Zhou}}, \bibinfo {author} {\bibfnamefont {K.}~\bibnamefont {Taguchi}},
  \bibinfo {author} {\bibfnamefont {Y.}~\bibnamefont {Kawaguchi}}, \bibinfo
  {author} {\bibfnamefont {Y.}~\bibnamefont {Tanaka}},\ and\ \bibinfo {author}
  {\bibfnamefont {K.~T.}\ \bibnamefont {Law}},\ }\href
  {https://doi.org/10.1038/s42005-019-0127-7} {\bibfield  {journal} {\bibinfo
  {journal} {Commun. Phys.}\ }\textbf {\bibinfo {volume} {2}},\ \bibinfo
  {pages} {26} (\bibinfo {year} {2019})}\BibitemShut {NoStop}%
\bibitem [{\citenamefont {McCann}(2006)}]{mccann}%
  \BibitemOpen
  \bibfield  {author} {\bibinfo {author} {\bibfnamefont {E.}~\bibnamefont
  {McCann}},\ }\href {https://doi.org/10.1103/PhysRevB.74.161403} {\bibfield
  {journal} {\bibinfo  {journal} {Phys. Rev. B}\ }\textbf {\bibinfo {volume}
  {74}},\ \bibinfo {pages} {161403(R)} (\bibinfo {year} {2006})}\BibitemShut
  {NoStop}%
\bibitem [{\citenamefont {Moulsdale}\ \emph {et~al.}(2020)\citenamefont
  {Moulsdale}, \citenamefont {Knothe},\ and\ \citenamefont
  {Fal'ko}}]{moulsdale}%
  \BibitemOpen
  \bibfield  {author} {\bibinfo {author} {\bibfnamefont {C.}~\bibnamefont
  {Moulsdale}}, \bibinfo {author} {\bibfnamefont {A.}~\bibnamefont {Knothe}},\
  and\ \bibinfo {author} {\bibfnamefont {V.}~\bibnamefont {Fal'ko}},\ }\href
  {https://doi.org/10.1103/PhysRevB.101.085118} {\bibfield  {journal} {\bibinfo
   {journal} {Phys. Rev. B}\ }\textbf {\bibinfo {volume} {101}},\ \bibinfo
  {pages} {085118} (\bibinfo {year} {2020})}\BibitemShut {NoStop}%
\bibitem [{\citenamefont {Battilomo}\ \emph {et~al.}(2019)\citenamefont
  {Battilomo}, \citenamefont {Scopigno},\ and\ \citenamefont
  {Ortix}}]{berrydipole}%
  \BibitemOpen
  \bibfield  {author} {\bibinfo {author} {\bibfnamefont {R.}~\bibnamefont
  {Battilomo}}, \bibinfo {author} {\bibfnamefont {N.}~\bibnamefont
  {Scopigno}},\ and\ \bibinfo {author} {\bibfnamefont {C.}~\bibnamefont
  {Ortix}},\ }\href {https://doi.org/10.1103/PhysRevLett.123.196403} {\bibfield
   {journal} {\bibinfo  {journal} {Phys. Rev. Lett.}\ }\textbf {\bibinfo
  {volume} {123}},\ \bibinfo {pages} {196403} (\bibinfo {year}
  {2019})}\BibitemShut {NoStop}%
\bibitem [{\citenamefont {Xu}\ \emph {et~al.}(2018)\citenamefont {Xu},
  \citenamefont {Ma}, \citenamefont {Shen}, \citenamefont {Fatemi},
  \citenamefont {Wu}, \citenamefont {Chang}, \citenamefont {Chang},
  \citenamefont {Valdivia}, \citenamefont {Chan}, \citenamefont {Gibson},
  \citenamefont {Zhou}, \citenamefont {Liu}, \citenamefont {Watanabe},
  \citenamefont {Taniguchi}, \citenamefont {Lin}, \citenamefont {Cava},
  \citenamefont {Fu}, \citenamefont {Gedik},\ and\ \citenamefont
  {Jarillo-Herrero}}]{berrywte2}%
  \BibitemOpen
  \bibfield  {author} {\bibinfo {author} {\bibfnamefont {S.-Y.}\ \bibnamefont
  {Xu}}, \bibinfo {author} {\bibfnamefont {Q.}~\bibnamefont {Ma}}, \bibinfo
  {author} {\bibfnamefont {H.}~\bibnamefont {Shen}}, \bibinfo {author}
  {\bibfnamefont {V.}~\bibnamefont {Fatemi}}, \bibinfo {author} {\bibfnamefont
  {S.}~\bibnamefont {Wu}}, \bibinfo {author} {\bibfnamefont {T.-R.}\
  \bibnamefont {Chang}}, \bibinfo {author} {\bibfnamefont {G.}~\bibnamefont
  {Chang}}, \bibinfo {author} {\bibfnamefont {A.~M.~M.}\ \bibnamefont
  {Valdivia}}, \bibinfo {author} {\bibfnamefont {C.-K.}\ \bibnamefont {Chan}},
  \bibinfo {author} {\bibfnamefont {Q.~D.}\ \bibnamefont {Gibson}}, \bibinfo
  {author} {\bibfnamefont {J.}~\bibnamefont {Zhou}}, \bibinfo {author}
  {\bibfnamefont {Z.}~\bibnamefont {Liu}}, \bibinfo {author} {\bibfnamefont
  {K.}~\bibnamefont {Watanabe}}, \bibinfo {author} {\bibfnamefont
  {T.}~\bibnamefont {Taniguchi}}, \bibinfo {author} {\bibfnamefont
  {H.}~\bibnamefont {Lin}}, \bibinfo {author} {\bibfnamefont {R.~J.}\
  \bibnamefont {Cava}}, \bibinfo {author} {\bibfnamefont {L.}~\bibnamefont
  {Fu}}, \bibinfo {author} {\bibfnamefont {N.}~\bibnamefont {Gedik}},\ and\
  \bibinfo {author} {\bibfnamefont {P.}~\bibnamefont {Jarillo-Herrero}},\
  }\href {https://doi.org/10.1038/s41567-018-0189-6} {\bibfield  {journal}
  {\bibinfo  {journal} {Nat. Phys.}\ }\textbf {\bibinfo {volume} {14}},\
  \bibinfo {pages} {900} (\bibinfo {year} {2018})}\BibitemShut {NoStop}%
\bibitem [{\citenamefont {Ma}\ \emph {et~al.}(2019)\citenamefont {Ma},
  \citenamefont {Xu}, \citenamefont {Shen}, \citenamefont {MacNeill},
  \citenamefont {Fatemi}, \citenamefont {Chang}, \citenamefont {{Mier
  Valdivia}}, \citenamefont {Wu}, \citenamefont {Du}, \citenamefont {Hsu},
  \citenamefont {Fang}, \citenamefont {Gibson}, \citenamefont {Watanabe},
  \citenamefont {Taniguchi}, \citenamefont {Cava}, \citenamefont {Kaxiras},
  \citenamefont {Lu}, \citenamefont {Lin}, \citenamefont {Fu}, \citenamefont
  {Gedik},\ and\ \citenamefont {Jarillo-Herrero}}]{berrywte2b}%
  \BibitemOpen
  \bibfield  {author} {\bibinfo {author} {\bibfnamefont {Q.}~\bibnamefont
  {Ma}}, \bibinfo {author} {\bibfnamefont {S.-Y.}\ \bibnamefont {Xu}}, \bibinfo
  {author} {\bibfnamefont {H.}~\bibnamefont {Shen}}, \bibinfo {author}
  {\bibfnamefont {D.}~\bibnamefont {MacNeill}}, \bibinfo {author}
  {\bibfnamefont {V.}~\bibnamefont {Fatemi}}, \bibinfo {author} {\bibfnamefont
  {T.-r.}\ \bibnamefont {Chang}}, \bibinfo {author} {\bibfnamefont {A.~M.}\
  \bibnamefont {{Mier Valdivia}}}, \bibinfo {author} {\bibfnamefont
  {S.}~\bibnamefont {Wu}}, \bibinfo {author} {\bibfnamefont {Z.}~\bibnamefont
  {Du}}, \bibinfo {author} {\bibfnamefont {C.-h.}\ \bibnamefont {Hsu}},
  \bibinfo {author} {\bibfnamefont {S.}~\bibnamefont {Fang}}, \bibinfo {author}
  {\bibfnamefont {Q.~D.}\ \bibnamefont {Gibson}}, \bibinfo {author}
  {\bibfnamefont {K.}~\bibnamefont {Watanabe}}, \bibinfo {author}
  {\bibfnamefont {T.}~\bibnamefont {Taniguchi}}, \bibinfo {author}
  {\bibfnamefont {R.~J.}\ \bibnamefont {Cava}}, \bibinfo {author}
  {\bibfnamefont {E.}~\bibnamefont {Kaxiras}}, \bibinfo {author} {\bibfnamefont
  {H.-Z.}\ \bibnamefont {Lu}}, \bibinfo {author} {\bibfnamefont
  {H.}~\bibnamefont {Lin}}, \bibinfo {author} {\bibfnamefont {L.}~\bibnamefont
  {Fu}}, \bibinfo {author} {\bibfnamefont {N.}~\bibnamefont {Gedik}},\ and\
  \bibinfo {author} {\bibfnamefont {P.}~\bibnamefont {Jarillo-Herrero}},\
  }\href {https://doi.org/10.1038/s41586-018-0807-6} {\bibfield  {journal}
  {\bibinfo  {journal} {Nature}\ }\textbf {\bibinfo {volume} {565}},\ \bibinfo
  {pages} {337} (\bibinfo {year} {2019})}\BibitemShut {NoStop}%
\bibitem [{\citenamefont {Kang}\ \emph {et~al.}(2019)\citenamefont {Kang},
  \citenamefont {Li}, \citenamefont {Sohn}, \citenamefont {Shan},\ and\
  \citenamefont {Mak}}]{kaifei}%
  \BibitemOpen
  \bibfield  {author} {\bibinfo {author} {\bibfnamefont {K.}~\bibnamefont
  {Kang}}, \bibinfo {author} {\bibfnamefont {T.}~\bibnamefont {Li}}, \bibinfo
  {author} {\bibfnamefont {E.}~\bibnamefont {Sohn}}, \bibinfo {author}
  {\bibfnamefont {J.}~\bibnamefont {Shan}},\ and\ \bibinfo {author}
  {\bibfnamefont {K.~F.}\ \bibnamefont {Mak}},\ }\href
  {https://doi.org/10.1038/s41563-019-0294-7} {\bibfield  {journal} {\bibinfo
  {journal} {Nat. Mater.}\ }\textbf {\bibinfo {volume} {18}},\ \bibinfo {pages}
  {324} (\bibinfo {year} {2019})}\BibitemShut {NoStop}%
\bibitem [{\citenamefont {Sodemann}\ and\ \citenamefont {Fu}(2015)}]{sodemann}%
  \BibitemOpen
  \bibfield  {author} {\bibinfo {author} {\bibfnamefont {I.}~\bibnamefont
  {Sodemann}}\ and\ \bibinfo {author} {\bibfnamefont {L.}~\bibnamefont {Fu}},\
  }\href {https://doi.org/10.1103/PhysRevLett.115.216806} {\bibfield  {journal}
  {\bibinfo  {journal} {Phys. Rev. Lett.}\ }\textbf {\bibinfo {volume} {115}},\
  \bibinfo {pages} {216806} (\bibinfo {year} {2015})}\BibitemShut {NoStop}%
\bibitem [{\citenamefont {Sharpe}\ \emph {et~al.}(2019)\citenamefont {Sharpe},
  \citenamefont {Fox}, \citenamefont {Barnard}, \citenamefont {Finney},
  \citenamefont {Watanabe}, \citenamefont {Taniguchi}, \citenamefont
  {Kastner},\ and\ \citenamefont {Goldhaber-Gordon}}]{dgg}%
  \BibitemOpen
  \bibfield  {author} {\bibinfo {author} {\bibfnamefont {A.~L.}\ \bibnamefont
  {Sharpe}}, \bibinfo {author} {\bibfnamefont {E.~J.}\ \bibnamefont {Fox}},
  \bibinfo {author} {\bibfnamefont {A.~W.}\ \bibnamefont {Barnard}}, \bibinfo
  {author} {\bibfnamefont {J.}~\bibnamefont {Finney}}, \bibinfo {author}
  {\bibfnamefont {K.}~\bibnamefont {Watanabe}}, \bibinfo {author}
  {\bibfnamefont {T.}~\bibnamefont {Taniguchi}}, \bibinfo {author}
  {\bibfnamefont {M.~A.}\ \bibnamefont {Kastner}},\ and\ \bibinfo {author}
  {\bibfnamefont {D.}~\bibnamefont {Goldhaber-Gordon}},\ }\href
  {https://doi.org/10.1126/science.aaw3780} {\bibfield  {journal} {\bibinfo
  {journal} {Science}\ }\textbf {\bibinfo {volume} {365}},\ \bibinfo {pages}
  {605} (\bibinfo {year} {2019})}\BibitemShut {NoStop}%
\bibitem [{\citenamefont {Sharpe}\ \emph {et~al.}()\citenamefont {Sharpe},
  \citenamefont {Fox}, \citenamefont {Barnard}, \citenamefont {Finney},
  \citenamefont {Watanabe}, \citenamefont {Taniguchi}, \citenamefont
  {Kastner},\ and\ \citenamefont {Goldhaber-Gordon}}]{orbitalferromagnetism}%
  \BibitemOpen
  \bibfield  {author} {\bibinfo {author} {\bibfnamefont {A.~L.}\ \bibnamefont
  {Sharpe}}, \bibinfo {author} {\bibfnamefont {E.~J.}\ \bibnamefont {Fox}},
  \bibinfo {author} {\bibfnamefont {A.~W.}\ \bibnamefont {Barnard}}, \bibinfo
  {author} {\bibfnamefont {J.}~\bibnamefont {Finney}}, \bibinfo {author}
  {\bibfnamefont {K.}~\bibnamefont {Watanabe}}, \bibinfo {author}
  {\bibfnamefont {T.}~\bibnamefont {Taniguchi}}, \bibinfo {author}
  {\bibfnamefont {M.~A.}\ \bibnamefont {Kastner}},\ and\ \bibinfo {author}
  {\bibfnamefont {D.}~\bibnamefont {Goldhaber-Gordon}},\ }\href
  {http://arxiv.org/abs/2102.04039} {\bibinfo  {journal} {arXiv:2102.04039}\
  }\BibitemShut {NoStop}%
\bibitem [{\citenamefont {Lu}\ \emph {et~al.}(2019)\citenamefont {Lu},
  \citenamefont {Stepanov}, \citenamefont {Yang}, \citenamefont {Xie},
  \citenamefont {Aamir}, \citenamefont {Das}, \citenamefont {Urgell},
  \citenamefont {Watanabe}, \citenamefont {Taniguchi}, \citenamefont {Zhang},
  \citenamefont {Bachtold}, \citenamefont {MacDonald},\ and\ \citenamefont
  {Efetov}}]{orbmagEfetov}%
  \BibitemOpen
\bibfield  {journal} {  }\bibfield  {author} {\bibinfo {author} {\bibfnamefont
  {X.}~\bibnamefont {Lu}}, \bibinfo {author} {\bibfnamefont {P.}~\bibnamefont
  {Stepanov}}, \bibinfo {author} {\bibfnamefont {W.}~\bibnamefont {Yang}},
  \bibinfo {author} {\bibfnamefont {M.}~\bibnamefont {Xie}}, \bibinfo {author}
  {\bibfnamefont {M.~A.}\ \bibnamefont {Aamir}}, \bibinfo {author}
  {\bibfnamefont {I.}~\bibnamefont {Das}}, \bibinfo {author} {\bibfnamefont
  {C.}~\bibnamefont {Urgell}}, \bibinfo {author} {\bibfnamefont
  {K.}~\bibnamefont {Watanabe}}, \bibinfo {author} {\bibfnamefont
  {T.}~\bibnamefont {Taniguchi}}, \bibinfo {author} {\bibfnamefont
  {G.}~\bibnamefont {Zhang}}, \bibinfo {author} {\bibfnamefont
  {A.}~\bibnamefont {Bachtold}}, \bibinfo {author} {\bibfnamefont {A.~H.}\
  \bibnamefont {MacDonald}},\ and\ \bibinfo {author} {\bibfnamefont {D.~K.}\
  \bibnamefont {Efetov}},\ }\href {https://doi.org/10.1038/s41586-019-1695-0}
  {\bibfield  {journal} {\bibinfo  {journal} {Nature}\ }\textbf {\bibinfo
  {volume} {574}},\ \bibinfo {pages} {653} (\bibinfo {year}
  {2019})}\BibitemShut {NoStop}%
\bibitem [{\citenamefont {He}\ \emph {et~al.}(2020)\citenamefont {He},
  \citenamefont {Goldhaber-Gordon},\ and\ \citenamefont {Law}}]{he}%
  \BibitemOpen
  \bibfield  {author} {\bibinfo {author} {\bibfnamefont {W.-Y.}\ \bibnamefont
  {He}}, \bibinfo {author} {\bibfnamefont {D.}~\bibnamefont
  {Goldhaber-Gordon}},\ and\ \bibinfo {author} {\bibfnamefont {K.~T.}\
  \bibnamefont {Law}},\ }\href {https://doi.org/10.1038/s41467-020-15473-9}
  {\bibfield  {journal} {\bibinfo  {journal} {Nat. Commun.}\ }\textbf {\bibinfo
  {volume} {11}},\ \bibinfo {pages} {1650} (\bibinfo {year}
  {2020})}\BibitemShut {NoStop}%
\bibitem [{\citenamefont {He}\ and\ \citenamefont {Law}()}]{he2}%
  \BibitemOpen
  \bibfield  {author} {\bibinfo {author} {\bibfnamefont {W.-Y.}\ \bibnamefont
  {He}}\ and\ \bibinfo {author} {\bibfnamefont {K.~T.}\ \bibnamefont {Law}},\
  }\href {http://arxiv.org/abs/2012.09896} {\bibinfo  {journal}
  {arXiv:2012.09896}\ }\BibitemShut {NoStop}%
\bibitem [{\citenamefont {Zhang}\ \emph {et~al.}()\citenamefont {Zhang},
  \citenamefont {Xiao}, \citenamefont {Zhou}, \citenamefont {Hu}, \citenamefont
  {Xie}, \citenamefont {Yan},\ and\ \citenamefont {Law}}]{strainTBG}%
  \BibitemOpen
\bibfield  {journal} {  }\bibfield  {author} {\bibinfo {author} {\bibfnamefont
  {C.-P.}\ \bibnamefont {Zhang}}, \bibinfo {author} {\bibfnamefont
  {J.}~\bibnamefont {Xiao}}, \bibinfo {author} {\bibfnamefont {B.~T.}\
  \bibnamefont {Zhou}}, \bibinfo {author} {\bibfnamefont {J.-X.}\ \bibnamefont
  {Hu}}, \bibinfo {author} {\bibfnamefont {Y.-M.}\ \bibnamefont {Xie}},
  \bibinfo {author} {\bibfnamefont {B.}~\bibnamefont {Yan}},\ and\ \bibinfo
  {author} {\bibfnamefont {K.~T.}\ \bibnamefont {Law}},\ }\href
  {http://arxiv.org/abs/2010.08333} {\bibinfo  {journal} {arXiv:2010.08333}\
  }\BibitemShut {NoStop}%
\bibitem [{\citenamefont {Tschirhart}\ \emph {et~al.}()\citenamefont
  {Tschirhart}, \citenamefont {Serlin}, \citenamefont {Polshyn}, \citenamefont
  {Shragai}, \citenamefont {Xia}, \citenamefont {Zhu}, \citenamefont {Zhang},
  \citenamefont {Watanabe}, \citenamefont {Taniguchi}, \citenamefont {Huber},\
  and\ \citenamefont {Young}}]{sotmoire}%
  \BibitemOpen
\bibfield  {journal} {  }\bibfield  {author} {\bibinfo {author} {\bibfnamefont
  {C.~L.}\ \bibnamefont {Tschirhart}}, \bibinfo {author} {\bibfnamefont
  {M.}~\bibnamefont {Serlin}}, \bibinfo {author} {\bibfnamefont
  {H.}~\bibnamefont {Polshyn}}, \bibinfo {author} {\bibfnamefont
  {A.}~\bibnamefont {Shragai}}, \bibinfo {author} {\bibfnamefont
  {Z.}~\bibnamefont {Xia}}, \bibinfo {author} {\bibfnamefont {J.}~\bibnamefont
  {Zhu}}, \bibinfo {author} {\bibfnamefont {Y.}~\bibnamefont {Zhang}}, \bibinfo
  {author} {\bibfnamefont {K.}~\bibnamefont {Watanabe}}, \bibinfo {author}
  {\bibfnamefont {T.}~\bibnamefont {Taniguchi}}, \bibinfo {author}
  {\bibfnamefont {M.~E.}\ \bibnamefont {Huber}},\ and\ \bibinfo {author}
  {\bibfnamefont {A.~F.}\ \bibnamefont {Young}},\ }\href
  {http://arxiv.org/abs/2006.08053} {\bibinfo  {journal} {arXiv:2006.08053}\
  }\BibitemShut {NoStop}%
\bibitem [{\citenamefont {Li}\ \emph {et~al.}(2020{\natexlab{b}})\citenamefont
  {Li}, \citenamefont {Zhang}, \citenamefont {Ren}, \citenamefont {Liu},
  \citenamefont {Dai},\ and\ \citenamefont {He}}]{moirestm}%
  \BibitemOpen
\bibfield  {journal} {  }\bibfield  {author} {\bibinfo {author} {\bibfnamefont
  {S.-Y.}\ \bibnamefont {Li}}, \bibinfo {author} {\bibfnamefont
  {Y.}~\bibnamefont {Zhang}}, \bibinfo {author} {\bibfnamefont {Y.-N.}\
  \bibnamefont {Ren}}, \bibinfo {author} {\bibfnamefont {J.}~\bibnamefont
  {Liu}}, \bibinfo {author} {\bibfnamefont {X.}~\bibnamefont {Dai}},\ and\
  \bibinfo {author} {\bibfnamefont {L.}~\bibnamefont {He}},\ }\href
  {https://doi.org/10.1103/PhysRevB.102.121406} {\bibfield  {journal} {\bibinfo
   {journal} {Phys. Rev. B}\ }\textbf {\bibinfo {volume} {102}},\ \bibinfo
  {pages} {121406(R)} (\bibinfo {year} {2020}{\natexlab{b}})}\BibitemShut
  {NoStop}%
\bibitem [{\citenamefont {Kuzmenko}\ \emph {et~al.}(2009)\citenamefont
  {Kuzmenko}, \citenamefont {Crassee}, \citenamefont {van~der Marel},
  \citenamefont {Blake},\ and\ \citenamefont {Novoselov}}]{hoppingparams}%
  \BibitemOpen
  \bibfield  {author} {\bibinfo {author} {\bibfnamefont {A.~B.}\ \bibnamefont
  {Kuzmenko}}, \bibinfo {author} {\bibfnamefont {I.}~\bibnamefont {Crassee}},
  \bibinfo {author} {\bibfnamefont {D.}~\bibnamefont {van~der Marel}}, \bibinfo
  {author} {\bibfnamefont {P.}~\bibnamefont {Blake}},\ and\ \bibinfo {author}
  {\bibfnamefont {K.~S.}\ \bibnamefont {Novoselov}},\ }\href
  {https://doi.org/10.1103/PhysRevB.80.165406} {\bibfield  {journal} {\bibinfo
  {journal} {Phys. Rev. B}\ }\textbf {\bibinfo {volume} {80}},\ \bibinfo
  {pages} {165406} (\bibinfo {year} {2009})}\BibitemShut {NoStop}%
\bibitem [{\citenamefont {Peres}\ \emph {et~al.}(2006)\citenamefont {Peres},
  \citenamefont {Guinea},\ and\ \citenamefont {{Castro Neto}}}]{peres}%
  \BibitemOpen
  \bibfield  {author} {\bibinfo {author} {\bibfnamefont {N.~M.~R.}\
  \bibnamefont {Peres}}, \bibinfo {author} {\bibfnamefont {F.}~\bibnamefont
  {Guinea}},\ and\ \bibinfo {author} {\bibfnamefont {A.~H.}\ \bibnamefont
  {{Castro Neto}}},\ }\href {https://doi.org/10.1103/PhysRevB.73.125411}
  {\bibfield  {journal} {\bibinfo  {journal} {Phys. Rev. B}\ }\textbf {\bibinfo
  {volume} {73}},\ \bibinfo {pages} {125411} (\bibinfo {year}
  {2006})}\BibitemShut {NoStop}%
\bibitem [{\citenamefont {Jung}\ and\ \citenamefont
  {MacDonald}(2014)}]{tbsign}%
  \BibitemOpen
  \bibfield  {author} {\bibinfo {author} {\bibfnamefont {J.}~\bibnamefont
  {Jung}}\ and\ \bibinfo {author} {\bibfnamefont {A.~H.}\ \bibnamefont
  {MacDonald}},\ }\href {https://doi.org/10.1103/PhysRevB.89.035405} {\bibfield
   {journal} {\bibinfo  {journal} {Phys. Rev. B}\ }\textbf {\bibinfo {volume}
  {89}},\ \bibinfo {pages} {035405} (\bibinfo {year} {2014})}\BibitemShut
  {NoStop}%
\bibitem [{\citenamefont {Joucken}\ \emph {et~al.}(2020)\citenamefont
  {Joucken}, \citenamefont {Ge}, \citenamefont {Quezada-L{\'{o}}pez},
  \citenamefont {Davenport}, \citenamefont {Watanabe}, \citenamefont
  {Taniguchi},\ and\ \citenamefont {Velasco}}]{warpingorientation}%
  \BibitemOpen
  \bibfield  {author} {\bibinfo {author} {\bibfnamefont {F.}~\bibnamefont
  {Joucken}}, \bibinfo {author} {\bibfnamefont {Z.}~\bibnamefont {Ge}},
  \bibinfo {author} {\bibfnamefont {E.~A.}\ \bibnamefont
  {Quezada-L{\'{o}}pez}}, \bibinfo {author} {\bibfnamefont {J.~L.}\
  \bibnamefont {Davenport}}, \bibinfo {author} {\bibfnamefont {K.}~\bibnamefont
  {Watanabe}}, \bibinfo {author} {\bibfnamefont {T.}~\bibnamefont
  {Taniguchi}},\ and\ \bibinfo {author} {\bibfnamefont {J.}~\bibnamefont
  {Velasco}},\ }\href {https://doi.org/10.1103/PhysRevB.101.161103} {\bibfield
  {journal} {\bibinfo  {journal} {Phys. Rev. B}\ }\textbf {\bibinfo {volume}
  {101}},\ \bibinfo {pages} {161103(R)} (\bibinfo {year} {2020})}\BibitemShut
  {NoStop}%
\bibitem [{\citenamefont {Mohiuddin}\ \emph {et~al.}(2009)\citenamefont
  {Mohiuddin}, \citenamefont {Lombardo}, \citenamefont {Nair}, \citenamefont
  {Bonetti}, \citenamefont {Savini}, \citenamefont {Jalil}, \citenamefont
  {Bonini}, \citenamefont {Basko}, \citenamefont {Galiotis}, \citenamefont
  {Marzari}, \citenamefont {Novoselov}, \citenamefont {Geim},\ and\
  \citenamefont {Ferrari}}]{gruneisen}%
  \BibitemOpen
  \bibfield  {author} {\bibinfo {author} {\bibfnamefont {T.~M.~G.}\
  \bibnamefont {Mohiuddin}}, \bibinfo {author} {\bibfnamefont {A.}~\bibnamefont
  {Lombardo}}, \bibinfo {author} {\bibfnamefont {R.~R.}\ \bibnamefont {Nair}},
  \bibinfo {author} {\bibfnamefont {A.}~\bibnamefont {Bonetti}}, \bibinfo
  {author} {\bibfnamefont {G.}~\bibnamefont {Savini}}, \bibinfo {author}
  {\bibfnamefont {R.}~\bibnamefont {Jalil}}, \bibinfo {author} {\bibfnamefont
  {N.}~\bibnamefont {Bonini}}, \bibinfo {author} {\bibfnamefont {D.~M.}\
  \bibnamefont {Basko}}, \bibinfo {author} {\bibfnamefont {C.}~\bibnamefont
  {Galiotis}}, \bibinfo {author} {\bibfnamefont {N.}~\bibnamefont {Marzari}},
  \bibinfo {author} {\bibfnamefont {K.~S.}\ \bibnamefont {Novoselov}}, \bibinfo
  {author} {\bibfnamefont {A.~K.}\ \bibnamefont {Geim}},\ and\ \bibinfo
  {author} {\bibfnamefont {A.~C.}\ \bibnamefont {Ferrari}},\ }\href
  {https://doi.org/10.1103/PhysRevB.79.205433} {\bibfield  {journal} {\bibinfo
  {journal} {Phys. Rev. B}\ }\textbf {\bibinfo {volume} {79}},\ \bibinfo
  {pages} {205433} (\bibinfo {year} {2009})}\BibitemShut {NoStop}%
\bibitem [{\citenamefont {Mariani}\ \emph {et~al.}(2012)\citenamefont
  {Mariani}, \citenamefont {Pearce},\ and\ \citenamefont {von
  Oppen}}]{gaugefields}%
  \BibitemOpen
  \bibfield  {author} {\bibinfo {author} {\bibfnamefont {E.}~\bibnamefont
  {Mariani}}, \bibinfo {author} {\bibfnamefont {A.~J.}\ \bibnamefont
  {Pearce}},\ and\ \bibinfo {author} {\bibfnamefont {F.}~\bibnamefont {von
  Oppen}},\ }\href {https://doi.org/10.1103/PhysRevB.86.165448} {\bibfield
  {journal} {\bibinfo  {journal} {Phys. Rev. B}\ }\textbf {\bibinfo {volume}
  {86}},\ \bibinfo {pages} {165448} (\bibinfo {year} {2012})}\BibitemShut
  {NoStop}%
\bibitem [{\citenamefont {Varlet}\ \emph {et~al.}(2014)\citenamefont {Varlet},
  \citenamefont {Bischoff}, \citenamefont {Simonet}, \citenamefont {Watanabe},
  \citenamefont {Taniguchi}, \citenamefont {Ihn}, \citenamefont {Ensslin},
  \citenamefont {Mucha-Kruczy{\'{n}}ski},\ and\ \citenamefont
  {Fal'ko}}]{interlayerasymmetry}%
  \BibitemOpen
  \bibfield  {author} {\bibinfo {author} {\bibfnamefont {A.}~\bibnamefont
  {Varlet}}, \bibinfo {author} {\bibfnamefont {D.}~\bibnamefont {Bischoff}},
  \bibinfo {author} {\bibfnamefont {P.}~\bibnamefont {Simonet}}, \bibinfo
  {author} {\bibfnamefont {K.}~\bibnamefont {Watanabe}}, \bibinfo {author}
  {\bibfnamefont {T.}~\bibnamefont {Taniguchi}}, \bibinfo {author}
  {\bibfnamefont {T.}~\bibnamefont {Ihn}}, \bibinfo {author} {\bibfnamefont
  {K.}~\bibnamefont {Ensslin}}, \bibinfo {author} {\bibfnamefont
  {M.}~\bibnamefont {Mucha-Kruczy{\'{n}}ski}},\ and\ \bibinfo {author}
  {\bibfnamefont {V.~I.}\ \bibnamefont {Fal'ko}},\ }\href
  {https://doi.org/10.1103/PhysRevLett.113.116602} {\bibfield  {journal}
  {\bibinfo  {journal} {Phys. Rev. Lett.}\ }\textbf {\bibinfo {volume} {113}},\
  \bibinfo {pages} {116602} (\bibinfo {year} {2014})}\BibitemShut {NoStop}%
\bibitem [{\citenamefont {Mucha-Kruczy{\'{n}}ski}\ \emph
  {et~al.}(2010)\citenamefont {Mucha-Kruczy{\'{n}}ski}, \citenamefont
  {McCann},\ and\ \citenamefont {Fal'ko}}]{blgtopicalreview}%
  \BibitemOpen
  \bibfield  {author} {\bibinfo {author} {\bibfnamefont {M.}~\bibnamefont
  {Mucha-Kruczy{\'{n}}ski}}, \bibinfo {author} {\bibfnamefont {E.}~\bibnamefont
  {McCann}},\ and\ \bibinfo {author} {\bibfnamefont {V.~I.}\ \bibnamefont
  {Fal'ko}},\ }\href {https://doi.org/10.1088/0268-1242/25/3/033001} {\bibfield
   {journal} {\bibinfo  {journal} {Semicond. Sci. Technol.}\ }\textbf {\bibinfo
  {volume} {25}},\ \bibinfo {pages} {033001} (\bibinfo {year}
  {2010})}\BibitemShut {NoStop}%
\bibitem [{\citenamefont {Slonczewski}\ and\ \citenamefont
  {Weiss}(1958)}]{Slonczewski}%
  \BibitemOpen
  \bibfield  {author} {\bibinfo {author} {\bibfnamefont {J.~C.}\ \bibnamefont
  {Slonczewski}}\ and\ \bibinfo {author} {\bibfnamefont {P.~R.}\ \bibnamefont
  {Weiss}},\ }\href {https://doi.org/10.1103/PhysRev.109.272} {\bibfield
  {journal} {\bibinfo  {journal} {Phys. Rev.}\ }\textbf {\bibinfo {volume}
  {109}},\ \bibinfo {pages} {272} (\bibinfo {year} {1958})}\BibitemShut
  {NoStop}%
\bibitem [{\citenamefont {McClure}(1957)}]{McClure}%
  \BibitemOpen
  \bibfield  {author} {\bibinfo {author} {\bibfnamefont {J.~W.}\ \bibnamefont
  {McClure}},\ }\href {https://doi.org/10.1103/PhysRev.108.612} {\bibfield
  {journal} {\bibinfo  {journal} {Phys. Rev.}\ }\textbf {\bibinfo {volume}
  {108}},\ \bibinfo {pages} {612} (\bibinfo {year} {1957})}\BibitemShut
  {NoStop}%
\bibitem [{\citenamefont {Pereira}\ \emph {et~al.}(2009)\citenamefont
  {Pereira}, \citenamefont {{Castro Neto}},\ and\ \citenamefont
  {Peres}}]{tbstrain}%
  \BibitemOpen
  \bibfield  {author} {\bibinfo {author} {\bibfnamefont {V.~M.}\ \bibnamefont
  {Pereira}}, \bibinfo {author} {\bibfnamefont {A.~H.}\ \bibnamefont {{Castro
  Neto}}},\ and\ \bibinfo {author} {\bibfnamefont {N.~M.~R.}\ \bibnamefont
  {Peres}},\ }\href {https://doi.org/10.1103/PhysRevB.80.045401} {\bibfield
  {journal} {\bibinfo  {journal} {Phys. Rev. B}\ }\textbf {\bibinfo {volume}
  {80}},\ \bibinfo {pages} {045401} (\bibinfo {year} {2009})}\BibitemShut
  {NoStop}%
\bibitem [{\citenamefont {Naumis}\ \emph {et~al.}(2017)\citenamefont {Naumis},
  \citenamefont {Barraza-Lopez}, \citenamefont {Oliva-Leyva},\ and\
  \citenamefont {Terrones}}]{strainreview}%
  \BibitemOpen
  \bibfield  {author} {\bibinfo {author} {\bibfnamefont {G.~G.}\ \bibnamefont
  {Naumis}}, \bibinfo {author} {\bibfnamefont {S.}~\bibnamefont
  {Barraza-Lopez}}, \bibinfo {author} {\bibfnamefont {M.}~\bibnamefont
  {Oliva-Leyva}},\ and\ \bibinfo {author} {\bibfnamefont {H.}~\bibnamefont
  {Terrones}},\ }\href {https://doi.org/10.1088/1361-6633/aa74ef} {\bibfield
  {journal} {\bibinfo  {journal} {Rep. Prog. Phys.}\ }\textbf {\bibinfo
  {volume} {80}},\ \bibinfo {pages} {096501} (\bibinfo {year}
  {2017})}\BibitemShut {NoStop}%
\bibitem [{\citenamefont {{Ramezani Masir}}\ \emph {et~al.}(2013)\citenamefont
  {{Ramezani Masir}}, \citenamefont {Moldovan},\ and\ \citenamefont
  {Peeters}}]{revisited}%
  \BibitemOpen
  \bibfield  {author} {\bibinfo {author} {\bibfnamefont {M.}~\bibnamefont
  {{Ramezani Masir}}}, \bibinfo {author} {\bibfnamefont {D.}~\bibnamefont
  {Moldovan}},\ and\ \bibinfo {author} {\bibfnamefont {F.}~\bibnamefont
  {Peeters}},\ }\href {https://doi.org/10.1016/j.ssc.2013.04.001} {\bibfield
  {journal} {\bibinfo  {journal} {Solid State Commun.}\ }\textbf {\bibinfo
  {volume} {175-176}},\ \bibinfo {pages} {76} (\bibinfo {year}
  {2013})}\BibitemShut {NoStop}%
\bibitem [{\citenamefont {Polyzos}\ \emph {et~al.}(2015)\citenamefont
  {Polyzos}, \citenamefont {Bianchi}, \citenamefont {Rizzi}, \citenamefont
  {Koukaras}, \citenamefont {Parthenios}, \citenamefont {Papagelis},
  \citenamefont {Sordan},\ and\ \citenamefont {Galiotis}}]{buckling}%
  \BibitemOpen
  \bibfield  {author} {\bibinfo {author} {\bibfnamefont {I.}~\bibnamefont
  {Polyzos}}, \bibinfo {author} {\bibfnamefont {M.}~\bibnamefont {Bianchi}},
  \bibinfo {author} {\bibfnamefont {L.}~\bibnamefont {Rizzi}}, \bibinfo
  {author} {\bibfnamefont {E.~N.}\ \bibnamefont {Koukaras}}, \bibinfo {author}
  {\bibfnamefont {J.}~\bibnamefont {Parthenios}}, \bibinfo {author}
  {\bibfnamefont {K.}~\bibnamefont {Papagelis}}, \bibinfo {author}
  {\bibfnamefont {R.}~\bibnamefont {Sordan}},\ and\ \bibinfo {author}
  {\bibfnamefont {C.}~\bibnamefont {Galiotis}},\ }\href
  {https://doi.org/10.1039/c5nr03072b} {\bibfield  {journal} {\bibinfo
  {journal} {Nanoscale}\ }\textbf {\bibinfo {volume} {7}},\ \bibinfo {pages}
  {13033} (\bibinfo {year} {2015})}\BibitemShut {NoStop}%
\bibitem [{\citenamefont {Frank}\ \emph {et~al.}(2010)\citenamefont {Frank},
  \citenamefont {Tsoukleri}, \citenamefont {Parthenios}, \citenamefont
  {Papagelis}, \citenamefont {Riaz}, \citenamefont {Jalil}, \citenamefont
  {Novoselov},\ and\ \citenamefont {Galiotis}}]{buckling3}%
  \BibitemOpen
  \bibfield  {author} {\bibinfo {author} {\bibfnamefont {O.}~\bibnamefont
  {Frank}}, \bibinfo {author} {\bibfnamefont {G.}~\bibnamefont {Tsoukleri}},
  \bibinfo {author} {\bibfnamefont {J.}~\bibnamefont {Parthenios}}, \bibinfo
  {author} {\bibfnamefont {K.}~\bibnamefont {Papagelis}}, \bibinfo {author}
  {\bibfnamefont {I.}~\bibnamefont {Riaz}}, \bibinfo {author} {\bibfnamefont
  {R.}~\bibnamefont {Jalil}}, \bibinfo {author} {\bibfnamefont {K.~S.}\
  \bibnamefont {Novoselov}},\ and\ \bibinfo {author} {\bibfnamefont
  {C.}~\bibnamefont {Galiotis}},\ }\href {https://doi.org/10.1021/nn100454w}
  {\bibfield  {journal} {\bibinfo  {journal} {ACS Nano}\ }\textbf {\bibinfo
  {volume} {4}},\ \bibinfo {pages} {3131} (\bibinfo {year} {2010})}\BibitemShut
  {NoStop}%
\bibitem [{\citenamefont {Xiao}\ \emph {et~al.}(2005)\citenamefont {Xiao},
  \citenamefont {Shi},\ and\ \citenamefont {Niu}}]{berrycorrection}%
  \BibitemOpen
  \bibfield  {author} {\bibinfo {author} {\bibfnamefont {D.}~\bibnamefont
  {Xiao}}, \bibinfo {author} {\bibfnamefont {J.}~\bibnamefont {Shi}},\ and\
  \bibinfo {author} {\bibfnamefont {Q.}~\bibnamefont {Niu}},\ }\href
  {https://doi.org/10.1103/PhysRevLett.95.137204} {\bibfield  {journal}
  {\bibinfo  {journal} {Phys. Rev. Lett.}\ }\textbf {\bibinfo {volume} {95}},\
  \bibinfo {pages} {137204} (\bibinfo {year} {2005})}\BibitemShut {NoStop}%
\bibitem [{\citenamefont {Thonhauser}\ \emph {et~al.}(2005)\citenamefont
  {Thonhauser}, \citenamefont {Ceresoli}, \citenamefont {Vanderbilt},\ and\
  \citenamefont {Resta}}]{thonhauser}%
  \BibitemOpen
  \bibfield  {author} {\bibinfo {author} {\bibfnamefont {T.}~\bibnamefont
  {Thonhauser}}, \bibinfo {author} {\bibfnamefont {D.}~\bibnamefont
  {Ceresoli}}, \bibinfo {author} {\bibfnamefont {D.}~\bibnamefont
  {Vanderbilt}},\ and\ \bibinfo {author} {\bibfnamefont {R.}~\bibnamefont
  {Resta}},\ }\href {https://doi.org/10.1103/PhysRevLett.95.137205} {\bibfield
  {journal} {\bibinfo  {journal} {Phys. Rev. Lett.}\ }\textbf {\bibinfo
  {volume} {95}},\ \bibinfo {pages} {137205} (\bibinfo {year}
  {2005})}\BibitemShut {NoStop}%
\bibitem [{\citenamefont {Shi}\ \emph {et~al.}(2007)\citenamefont {Shi},
  \citenamefont {Vignale}, \citenamefont {Xiao},\ and\ \citenamefont
  {Niu}}]{orbitalmagderivation}%
  \BibitemOpen
  \bibfield  {author} {\bibinfo {author} {\bibfnamefont {J.}~\bibnamefont
  {Shi}}, \bibinfo {author} {\bibfnamefont {G.}~\bibnamefont {Vignale}},
  \bibinfo {author} {\bibfnamefont {D.}~\bibnamefont {Xiao}},\ and\ \bibinfo
  {author} {\bibfnamefont {Q.}~\bibnamefont {Niu}},\ }\href
  {https://doi.org/10.1103/PhysRevLett.99.197202} {\bibfield  {journal}
  {\bibinfo  {journal} {Phys. Rev. Lett.}\ }\textbf {\bibinfo {volume} {99}},\
  \bibinfo {pages} {197202} (\bibinfo {year} {2007})}\BibitemShut {NoStop}%
\bibitem [{\citenamefont {Ashcroft}\ and\ \citenamefont
  {Mermin}(1976)}]{ashcroftmermin}%
  \BibitemOpen
  \bibfield  {author} {\bibinfo {author} {\bibfnamefont {N.}~\bibnamefont
  {Ashcroft}}\ and\ \bibinfo {author} {\bibfnamefont {D.}~\bibnamefont
  {Mermin}},\ }\href@noop {} {\emph {\bibinfo {title} {{Solid State
  Physics}}}}\ (\bibinfo  {publisher} {Saunders College Publishing},\ \bibinfo
  {address} {Fort Worth},\ \bibinfo {year} {1976})\BibitemShut {NoStop}%
\bibitem [{\citenamefont {Nam}\ \emph {et~al.}(2017)\citenamefont {Nam},
  \citenamefont {Ki}, \citenamefont {Soler-Delgado},\ and\ \citenamefont
  {Morpurgo}}]{relaxationtime1}%
  \BibitemOpen
  \bibfield  {author} {\bibinfo {author} {\bibfnamefont {Y.}~\bibnamefont
  {Nam}}, \bibinfo {author} {\bibfnamefont {D.-K.}\ \bibnamefont {Ki}},
  \bibinfo {author} {\bibfnamefont {D.}~\bibnamefont {Soler-Delgado}},\ and\
  \bibinfo {author} {\bibfnamefont {A.~F.}\ \bibnamefont {Morpurgo}},\ }\href
  {https://doi.org/10.1038/nphys4218} {\bibfield  {journal} {\bibinfo
  {journal} {Nat. Phys.}\ }\textbf {\bibinfo {volume} {13}},\ \bibinfo {pages}
  {1207} (\bibinfo {year} {2017})}\BibitemShut {NoStop}%
\bibitem [{\citenamefont {Hwang}\ and\ \citenamefont {{Das
  Sarma}}(2008)}]{relaxationtime2}%
  \BibitemOpen
  \bibfield  {author} {\bibinfo {author} {\bibfnamefont {E.~H.}\ \bibnamefont
  {Hwang}}\ and\ \bibinfo {author} {\bibfnamefont {S.}~\bibnamefont {{Das
  Sarma}}},\ }\href {https://doi.org/10.1103/PhysRevB.77.195412} {\bibfield
  {journal} {\bibinfo  {journal} {Phys. Rev. B}\ }\textbf {\bibinfo {volume}
  {77}},\ \bibinfo {pages} {195412} (\bibinfo {year} {2008})}\BibitemShut
  {NoStop}%
\bibitem [{\citenamefont {Wagner}\ \emph {et~al.}(2020)\citenamefont {Wagner},
  \citenamefont {Nguyen},\ and\ \citenamefont {Simon}}]{relaxationtime3}%
  \BibitemOpen
  \bibfield  {author} {\bibinfo {author} {\bibfnamefont {G.}~\bibnamefont
  {Wagner}}, \bibinfo {author} {\bibfnamefont {D.~X.}\ \bibnamefont {Nguyen}},\
  and\ \bibinfo {author} {\bibfnamefont {S.~H.}\ \bibnamefont {Simon}},\ }\href
  {https://doi.org/10.1103/PhysRevLett.124.026601} {\bibfield  {journal}
  {\bibinfo  {journal} {Phys. Rev. Lett.}\ }\textbf {\bibinfo {volume} {124}},\
  \bibinfo {pages} {026601} (\bibinfo {year} {2020})}\BibitemShut {NoStop}%
\bibitem [{\citenamefont {Bhowal}\ and\ \citenamefont {Satpathy}(2020)}]{nbx2}%
  \BibitemOpen
  \bibfield  {author} {\bibinfo {author} {\bibfnamefont {S.}~\bibnamefont
  {Bhowal}}\ and\ \bibinfo {author} {\bibfnamefont {S.}~\bibnamefont
  {Satpathy}},\ }\href {https://doi.org/10.1103/physrevb.102.201403} {\bibfield
   {journal} {\bibinfo  {journal} {Phys. Rev. B}\ }\textbf {\bibinfo {volume}
  {102}},\ \bibinfo {pages} {201403(R)} (\bibinfo {year} {2020})}\BibitemShut
  {NoStop}%
\bibitem [{\citenamefont {Wang}\ \emph {et~al.}(2019)\citenamefont {Wang},
  \citenamefont {Zihlmann}, \citenamefont {Baumgartner}, \citenamefont
  {Overbeck}, \citenamefont {Watanabe}, \citenamefont {Taniguchi},
  \citenamefont {Makk},\ and\ \citenamefont {Sch{\"{o}}nenberger}}]{lujun}%
  \BibitemOpen
  \bibfield  {author} {\bibinfo {author} {\bibfnamefont {L.}~\bibnamefont
  {Wang}}, \bibinfo {author} {\bibfnamefont {S.}~\bibnamefont {Zihlmann}},
  \bibinfo {author} {\bibfnamefont {A.}~\bibnamefont {Baumgartner}}, \bibinfo
  {author} {\bibfnamefont {J.}~\bibnamefont {Overbeck}}, \bibinfo {author}
  {\bibfnamefont {K.}~\bibnamefont {Watanabe}}, \bibinfo {author}
  {\bibfnamefont {T.}~\bibnamefont {Taniguchi}}, \bibinfo {author}
  {\bibfnamefont {P.}~\bibnamefont {Makk}},\ and\ \bibinfo {author}
  {\bibfnamefont {C.}~\bibnamefont {Sch{\"{o}}nenberger}},\ }\href
  {https://doi.org/10.1021/acs.nanolett.9b01491} {\bibfield  {journal}
  {\bibinfo  {journal} {Nano Lett.}\ }\textbf {\bibinfo {volume} {19}},\
  \bibinfo {pages} {4097} (\bibinfo {year} {2019})}\BibitemShut {NoStop}%
\bibitem [{\citenamefont {Wang}\ \emph {et~al.}(2020)\citenamefont {Wang},
  \citenamefont {Makk}, \citenamefont {Zihlmann}, \citenamefont {Baumgartner},
  \citenamefont {Indolese}, \citenamefont {Watanabe}, \citenamefont
  {Taniguchi},\ and\ \citenamefont {Sch{\"{o}}nenberger}}]{lujun2}%
  \BibitemOpen
  \bibfield  {author} {\bibinfo {author} {\bibfnamefont {L.}~\bibnamefont
  {Wang}}, \bibinfo {author} {\bibfnamefont {P.}~\bibnamefont {Makk}}, \bibinfo
  {author} {\bibfnamefont {S.}~\bibnamefont {Zihlmann}}, \bibinfo {author}
  {\bibfnamefont {A.}~\bibnamefont {Baumgartner}}, \bibinfo {author}
  {\bibfnamefont {D.~I.}\ \bibnamefont {Indolese}}, \bibinfo {author}
  {\bibfnamefont {K.}~\bibnamefont {Watanabe}}, \bibinfo {author}
  {\bibfnamefont {T.}~\bibnamefont {Taniguchi}},\ and\ \bibinfo {author}
  {\bibfnamefont {C.}~\bibnamefont {Sch{\"{o}}nenberger}},\ }\href
  {https://doi.org/10.1103/PhysRevLett.124.157701} {\bibfield  {journal}
  {\bibinfo  {journal} {Phys. Rev. Lett.}\ }\textbf {\bibinfo {volume} {124}},\
  \bibinfo {pages} {157701} (\bibinfo {year} {2020})}\BibitemShut {NoStop}%
\bibitem [{\citenamefont {Wang}\ \emph {et~al.}()\citenamefont {Wang},
  \citenamefont {Baumgartner}, \citenamefont {Makk}, \citenamefont {Zihlmann},
  \citenamefont {Varghese}, \citenamefont {Indolese}, \citenamefont {Watanabe},
  \citenamefont {Taniguchi},\ and\ \citenamefont
  {Sch{\"{o}}nenberger}}]{lujun3}%
  \BibitemOpen
  \bibfield  {author} {\bibinfo {author} {\bibfnamefont {L.}~\bibnamefont
  {Wang}}, \bibinfo {author} {\bibfnamefont {A.}~\bibnamefont {Baumgartner}},
  \bibinfo {author} {\bibfnamefont {P.}~\bibnamefont {Makk}}, \bibinfo {author}
  {\bibfnamefont {S.}~\bibnamefont {Zihlmann}}, \bibinfo {author}
  {\bibfnamefont {B.~S.}\ \bibnamefont {Varghese}}, \bibinfo {author}
  {\bibfnamefont {D.~I.}\ \bibnamefont {Indolese}}, \bibinfo {author}
  {\bibfnamefont {K.}~\bibnamefont {Watanabe}}, \bibinfo {author}
  {\bibfnamefont {T.}~\bibnamefont {Taniguchi}},\ and\ \bibinfo {author}
  {\bibfnamefont {C.}~\bibnamefont {Sch{\"{o}}nenberger}},\ }\href
  {http://arxiv.org/abs/2009.03035} {\bibinfo  {journal} {arXiv:2009.03035}\
  }\BibitemShut {NoStop}%
\bibitem [{\citenamefont {Vasyukov}\ \emph {et~al.}(2013)\citenamefont
  {Vasyukov}, \citenamefont {Anahory}, \citenamefont {Embon}, \citenamefont
  {Halbertal}, \citenamefont {Cuppens}, \citenamefont {Neeman}, \citenamefont
  {Finkler}, \citenamefont {Segev}, \citenamefont {Myasoedov}, \citenamefont
  {Rappaport}, \citenamefont {Huber},\ and\ \citenamefont {Zeldov}}]{sot}%
  \BibitemOpen
\bibfield  {journal} {  }\bibfield  {author} {\bibinfo {author} {\bibfnamefont
  {D.}~\bibnamefont {Vasyukov}}, \bibinfo {author} {\bibfnamefont
  {Y.}~\bibnamefont {Anahory}}, \bibinfo {author} {\bibfnamefont
  {L.}~\bibnamefont {Embon}}, \bibinfo {author} {\bibfnamefont
  {D.}~\bibnamefont {Halbertal}}, \bibinfo {author} {\bibfnamefont
  {J.}~\bibnamefont {Cuppens}}, \bibinfo {author} {\bibfnamefont
  {L.}~\bibnamefont {Neeman}}, \bibinfo {author} {\bibfnamefont
  {A.}~\bibnamefont {Finkler}}, \bibinfo {author} {\bibfnamefont
  {Y.}~\bibnamefont {Segev}}, \bibinfo {author} {\bibfnamefont
  {Y.}~\bibnamefont {Myasoedov}}, \bibinfo {author} {\bibfnamefont {M.~L.}\
  \bibnamefont {Rappaport}}, \bibinfo {author} {\bibfnamefont {M.~E.}\
  \bibnamefont {Huber}},\ and\ \bibinfo {author} {\bibfnamefont
  {E.}~\bibnamefont {Zeldov}},\ }\href {https://doi.org/10.1038/nnano.2013.169}
  {\bibfield  {journal} {\bibinfo  {journal} {Nat. Nanotechnol.}\ }\textbf
  {\bibinfo {volume} {8}},\ \bibinfo {pages} {639} (\bibinfo {year}
  {2013})}\BibitemShut {NoStop}%
\bibitem [{\citenamefont {Kirtley}(2010)}]{kirtleyreview}%
  \BibitemOpen
  \bibfield  {author} {\bibinfo {author} {\bibfnamefont {J.~R.}\ \bibnamefont
  {Kirtley}},\ }\href {https://doi.org/10.1088/0034-4885/73/12/126501}
  {\bibfield  {journal} {\bibinfo  {journal} {Rep. Prog. Phys.}\ }\textbf
  {\bibinfo {volume} {73}},\ \bibinfo {pages} {126501} (\bibinfo {year}
  {2010})}\BibitemShut {NoStop}%
\bibitem [{\citenamefont {Casola}\ \emph {et~al.}(2018)\citenamefont {Casola},
  \citenamefont {{Van Der Sar}},\ and\ \citenamefont {Yacoby}}]{nvreview}%
  \BibitemOpen
  \bibfield  {author} {\bibinfo {author} {\bibfnamefont {F.}~\bibnamefont
  {Casola}}, \bibinfo {author} {\bibfnamefont {T.}~\bibnamefont {{Van Der
  Sar}}},\ and\ \bibinfo {author} {\bibfnamefont {A.}~\bibnamefont {Yacoby}},\
  }\href {https://doi.org/10.1038/natrevmats.2017.88} {\bibfield  {journal}
  {\bibinfo  {journal} {Nat. Rev. Mater.}\ }\textbf {\bibinfo {volume} {3}},\
  \bibinfo {pages} {17088} (\bibinfo {year} {2018})}\BibitemShut {NoStop}%
\bibitem [{\citenamefont {Thiel}\ \emph {et~al.}(2019)\citenamefont {Thiel},
  \citenamefont {Wang}, \citenamefont {Tschudin}, \citenamefont {Rohner},
  \citenamefont {Guti{\'{e}}rrez-Lezama}, \citenamefont {Ubrig}, \citenamefont
  {Gibertini}, \citenamefont {Giannini}, \citenamefont {Morpurgo},\ and\
  \citenamefont {Maletinsky}}]{thiel}%
  \BibitemOpen
  \bibfield  {author} {\bibinfo {author} {\bibfnamefont {L.}~\bibnamefont
  {Thiel}}, \bibinfo {author} {\bibfnamefont {Z.}~\bibnamefont {Wang}},
  \bibinfo {author} {\bibfnamefont {M.~A.}\ \bibnamefont {Tschudin}}, \bibinfo
  {author} {\bibfnamefont {D.}~\bibnamefont {Rohner}}, \bibinfo {author}
  {\bibfnamefont {I.}~\bibnamefont {Guti{\'{e}}rrez-Lezama}}, \bibinfo {author}
  {\bibfnamefont {N.}~\bibnamefont {Ubrig}}, \bibinfo {author} {\bibfnamefont
  {M.}~\bibnamefont {Gibertini}}, \bibinfo {author} {\bibfnamefont
  {E.}~\bibnamefont {Giannini}}, \bibinfo {author} {\bibfnamefont {A.~F.}\
  \bibnamefont {Morpurgo}},\ and\ \bibinfo {author} {\bibfnamefont
  {P.}~\bibnamefont {Maletinsky}},\ }\href
  {https://doi.org/10.1126/science.aav6926} {\bibfield  {journal} {\bibinfo
  {journal} {Science}\ }\textbf {\bibinfo {volume} {364}},\ \bibinfo {pages}
  {973} (\bibinfo {year} {2019})}\BibitemShut {NoStop}%
\bibitem [{\citenamefont {Knothe}\ and\ \citenamefont
  {Fal'ko}(2018)}]{twobandmoment}%
  \BibitemOpen
  \bibfield  {author} {\bibinfo {author} {\bibfnamefont {A.}~\bibnamefont
  {Knothe}}\ and\ \bibinfo {author} {\bibfnamefont {V.}~\bibnamefont
  {Fal'ko}},\ }\href {https://doi.org/10.1103/PhysRevB.98.155435} {\bibfield
  {journal} {\bibinfo  {journal} {Phys. Rev. B}\ }\textbf {\bibinfo {volume}
  {98}},\ \bibinfo {pages} {155435} (\bibinfo {year} {2018})}\BibitemShut
  {NoStop}%
\bibitem [{\citenamefont {Park}(2018)}]{twobandmoment2}%
  \BibitemOpen
  \bibfield  {author} {\bibinfo {author} {\bibfnamefont {C.-S.}\ \bibnamefont
  {Park}},\ }\href {https://doi.org/10.1016/j.physleta.2017.10.044} {\bibfield
  {journal} {\bibinfo  {journal} {Phys. Lett. A}\ }\textbf {\bibinfo {volume}
  {382}},\ \bibinfo {pages} {121} (\bibinfo {year} {2018})}\BibitemShut
  {NoStop}%
\end{thebibliography}%

\end{document}